\documentclass{article}
\usepackage[T1]{fontenc}
\usepackage[utf8]{inputenc}

\usepackage[english]{babel}
\usepackage{csquotes}
\usepackage{authblk}

\usepackage{lscape}
\usepackage{amsmath}
\usepackage{mathtools}
\usepackage{amsfonts}
\usepackage{amssymb}
\usepackage{algorithm}
\usepackage{algpseudocode}
\usepackage{enumerate}
\usepackage{physics}
\usepackage{hyperref}
\usepackage{xcolor}

\usepackage{multirow}

\usepackage{comment}
\usepackage{standalone}
\usepackage{svg}
\usepackage{tikz}
\usepackage{authblk}
\usepackage{bbm}

\svgsetup{inkscapelatex=false}
\graphicspath{{Illustrations/}}
\usetikzlibrary{positioning}

\newcommand{\argmax}{\operatornamewithlimits{arg\,max}}
\newcommand{\argmin}{\operatornamewithlimits{arg\,min}}

\usepackage{natbib}
\defcitealias{aera_apa_ncme_2014}{American Educational Research Association et al., 2014}

\usepackage{float}
\usepackage[toc,page]{appendix}
\usepackage[onehalfspacing]{setspace}

\usepackage{adjustbox}
\usepackage{rotating}

\usepackage{booktabs,dcolumn,caption}

\usepackage{subcaption}

\providecommand{\keywords}[1]
{
  \small	
  \textit{Keywords:} #1
}

\makeatletter
\newcommand*\rel@kern[1]{\kern#1\dimexpr\macc@kerna}
\newcommand*\widebar[1]{%
  \begingroup
  \def\mathaccent##1##2{%
    \rel@kern{0.8}%
    \overline{\rel@kern{-0.8}\macc@nucleus\rel@kern{0.2}}%
    \rel@kern{-0.2}%
  }%
  \macc@depth\@ne
  \let\math@bgroup\@empty \let\math@egroup\macc@set@skewchar
  \mathsurround\z@ \frozen@everymath{\mathgroup\macc@group\relax}%
  \macc@set@skewchar\relax
  \let\mathaccentV\macc@nested@a
  \macc@nested@a\relax111{#1}%
  \endgroup
}
\makeatother

\newcolumntype{L}[1]{>{\raggedright\let\newline\\\arraybackslash\hspace{0pt}}p{#1}}
\newcolumntype{C}[1]{>{\centering\let\newline\\\arraybackslash\hspace{0pt}}p{#1}}
\newcolumntype{R}[1]{>{\raggedleft\let\newline\\\arraybackslash\hspace{0pt}}p{#1}}

\title{A Latent Variable Model with Change-Points and Its Application to Time Pressure Effects in Educational Assessment}

\author[1]{Gabriel Wallin}
\author[2]{Yunxiao Chen}
\author[3]{Yi-Hsuan Lee}
\author[4]{Xiaoou Li}

\affil[1]{School of Mathematical Sciences, Lancaster University}
\affil[2]{Department of Statistics, London School of Economics and Political Science}
\affil[3]{Independent Researcher}
\affil[4]{School of Statistics, University of Minnesota}

\date{}

\begin{document}

\maketitle
		
		\begin{abstract}
Educational assessments are valuable tools for measuring student knowledge and skills, but their validity can be compromised when test takers exhibit changes in response behavior due to factors such as time pressure. To address this issue, we introduce a novel latent factor model with change-points for item response data, designed to detect and account for individual-level shifts in response patterns during testing. This model extends traditional Item Response Theory (IRT) by incorporating person-specific change-points, which enables simultaneous estimation of item parameters, person latent traits, and the location of behavioral changes. We evaluate the proposed model through extensive simulation studies, which demonstrate its ability to accurately recover item parameters, change-point locations, and individual ability estimates under various conditions. Our findings show that accounting for change-points significantly reduces bias in ability estimates, particularly for respondents affected by time pressure. Application of the model to two real-world educational testing datasets reveals distinct patterns of change-point occurrence between high-stakes and lower-stakes tests, providing insights into how test-taking behavior evolves during the tests. This approach offers a more nuanced understanding of test-taking dynamics, with important implications for test design, scoring, and interpretation. 
		\end{abstract}


	

\section{Introduction}

Educational assessments play an important role for measuring student knowledge, skills, and growth in various academic domains. These tests are also used in educational decision-making, from informing instructional practices to high-stakes determinations such as college admissions. Given their importance, ensuring the validity and fairness of test score interpretations is paramount \citepalias[see Chapter 3,][]{aera_apa_ncme_2014}. However, these measurements can be compromised when test takers exhibit changes in response behavior during the test due to, for example, time-pressure, fatigue or disengagement \citep{boughton2007hybrid, goegebeur2008speeded, wise2006application, wang2015mixture}. 

In standardized educational testing, respondents are given a set of questions, referred to as items, constructed to measure the respondent's ability levels in some well-defined domains. Multiple-choice items are typically used, with one correct response category per item. In the educational testing literature, such items are often referred to as binary items since the responses are scored as either correct or incorrect. 
A common assumption when analysing such testing data is that every respondent is giving every item full attention \citep{bolt2002item, schnipke1997modeling, wise2005response}. In that way, the item responses can be assumed to only reflect the ability level of the respondent (plus some measurement error), given that the items are well-constructed. The probability of a correct answer therefore monotonically increases as the latent trait level increases \citep{wang2015mixture}. At the same time, standardized tests typically impose a time limit, terminating the test when the allotted time expires regardless of completion status. Due to time pressure, the imposed time limit can lead to a behavioral change, where the respondent responds without processing the item content \citep{wise2005response}. It is commonly known as test speededness in the educational testing literature. This change can be seen as a contamination of the signal of a latent variable: The respondent's latent ability is the primary factor affecting the item responses up until the change, but thereafter, the responses start getting affected by time pressure effects as well. Post-change responses do therefore not accurately reflect the respondent's true ability, which could lead to a potential misinterpretation of their performance. 

Item response theory (IRT) models have been extensively used in the development, assessment, and scoring of educational tests \citep{chen2021item, embretson2013item}. They provide a probabilistic framework that relates observed item responses to latent traits. IRT is the engine behind several educational testing innovations, such as (i) computerized adaptive testing \citep{van2000computerized}, which dynamically adjusts item difficulty to match a test-taker's ability in real time, (ii) multistage testing \citep{yan2016computerized}, which adjusts difficulty at predefined stages based on performance, (iii) automated test assembly \citep{linden2005linear}, which uses algorithms to create assessments that meet specific statistical and content requirements, and (iv) personalized learning systems \citep{chen2005personalized} which customize instruction, content, and pacing to individual student needs and abilities.  At the core of all these frameworks is an IRT model.

Traditional IRT models assume that all respondents give full attention to all items throughout the test, disregarding potential time-pressure effects or changes in problem-solving strategies \citep{van2007hierarchical}. In practice, test-taker behavior often deviates from this idealized assumption, particularly in timed testing scenarios. As test takers progress through an examination, factors such as fatigue, time constraints, or disengagement can lead to shifts in response patterns (Wise \& DeMars, 2006). One well-documented phenomenon is  ``rapid guessing'', where respondents switch from careful problem-solving to quick, often random, responding as time runs out \citep{schnipke1997modeling}. This behavior introduces construct-irrelevant variance into test scores, potentially biasing ability estimates and threatening the validity of inferences drawn from these assessments \citep{bolt2002item}.

The detection and modeling of such behavioral changes during testing have received growing attention in psychometric research. Early approaches to identifying aberrant response patterns, such as person-fit statistics \citep{meijer2001methodology}, provided global measures of response consistency but lacked the ability to pinpoint where changes in behavior occur. Other work has explored IRT models that allow for shifts in item parameters at a fixed point in the test, common to all test-takers \citep{bolt2002item, yamamoto1997modeling}. However, these models do not account for individual differences in the onset of behavioral changes, which may vary due to factors such as cognitive fatigue, time management skills, or the interaction between ability and item difficulty.

To address these limitations, we turn to the literature on change-point detection. Change-point methods have been widely applied to detect abrupt shifts in data generating processes across various fields. These include, but are not limited to, bioinformatics \citep{olshen2004circular, futschik2014multiscale}, climatology \citep{reeves2007review}, distributed sensor networks \citep{yang2024communication}, medicine \citep{bosc2003automatic, staudacher2005new}, entertainment \citep{rybach2009audio, radke2005image} and finance \citep{kim2005structural, andreou2002detecting}. We choose not to distinguish between online and offline settings, but instead point out that most of the developed change-point methods focus on either detecting a single change \citep[e.g.,][]{gombay1990asymptotic, hinkley1970inference, hinkley1971inference, hawkins1977testing, james1987tests, sen1975tests, worsley1979likelihood}, or multiple changes in a single data stream \citep[e.g.,][among many others]{romano2022detecting, killick2012optimal, fryzlewicz2014wild, fearnhead2019changepoint, yao1988estimating, zou2020consistent, niu2016multiple, maidstone2017optimal, fryzlewicz2018tail}. Of particular relevance to our work is the literature on change-point detection for multiple data streams \citep{chan2017optimal, chen2015graph, mei2010efficient, xie2013sequential, fellouris2016second}, \textcolor{black}{where in our context, each respondent's sequence of item responses represents a separate data stream}. However, most literature on multi-stream change-point detection focuses on detection problems with a shared change-point among the data streams. Exceptions include \cite{chen2023compound}, \cite{chen2022item}, and \cite{lu2022optimal}. It is noteworthy that in our setting, we require a method capable of detecting changes for each data stream at potentially different locations. Additionally, existing change-point methods do not take the complex interdependencies in IRT models into account. In IRT, item parameters and person abilities are simultaneously estimated, creating a multidimensional problem where changes in one component can affect all others. This interdependence requires the development of new methodologies that can simultaneously estimate item parameters, person abilities, and individual-specific change-points within the IRT framework.

The present research addresses this challenge by proposing a new class of IRT models that allow for individual-level change-points in response behavior. Specifically, we extend the standard two-parameter logistic (2-PL) IRT model \citep{birnbaum1968some} to allow item-specific parameters that carry an application-related interpretation to change after an item that varies by person. The location of the change-point is treated as a latent random variable, enabling us to estimate its distribution in the test-taker population. We develop an empirical Bayes estimation approach based on the marginal maximum likelihood function to simultaneously estimate item parameters, person abilities, and change-point locations. 

Our work contributes to an emerging line of psychometric research on learning and behavioral dynamics in assessment. Our contribution adds to a series of recent studies that have explored extensions of IRT models to capture within-person changes from aberrant behavior due to e.g., time presure, fatigue, or cheating. \cite{wang2015mixture} proposes a mixture hierarchical model that uses both item responses and response time to model effects of changes in response style, \cite{wang2018detecting} adjusts the model in \cite{wang2015mixture} to include item-level response-style effect parameters, and \cite{wang2018two} combines a mixture modeling approach with a residual-based outlier detection method to distinguish normal behavior from aberrant behavior. Similar to our approach, \cite{shao2016detection} formulates the person-level change as a change-point detection problem, but assumes all of the other model parameters to be known. Our proposal contributes to this stream of research by allowing for qualitative shifts in response behavior at a respondent-unique position while simultaneously estimating all the model parameters. This moves beyond static conceptions of ability and response behavior to provide a richer understanding of test-taker performance. 

The proposed methodology offers several potential benefits for educational measurement:
\begin{enumerate}
\item Enhanced understanding of test-taking behavior: By modeling individual-level change-points, we can gain insights into how response patterns evolve over the course of a test and how this may vary across test takers. 

\item Improved measurement accuracy: Accounting for behavioral shifts can lead to more accurate estimates of test-taker abilities, particularly for those who exhibit rapid guessing or other changes in response strategy.

\item Test design implications: Identifying patterns in change-point locations can inform decisions about test length, time limits, and item ordering to minimize construct-irrelevant variance.

\item Fairness considerations: Detecting and accounting for behavioral shifts can enhance the validity and fairness of test score interpretations, particularly in high-stakes testing contexts.
\end{enumerate}

To demonstrate the utility of our approach, we apply the proposed change-point IRT model to response data from two educational tests. These analyses provide new insights into how response behavior changes over the course of testing and its impact on the measurement of examinees' skills. We also conduct a simulation study to evaluate the model's performance under various conditions.

The remainder of this paper is organized as follows: Section 2 presents the proposed change-point IRT model in detail. Section 3 presents model inference, and Section 4 describes model generalizations. In Section 5, we present a simulation study to evaluate the model's performance. Section 6 provides an empirical analysis of two real educational tests using the proposed methodology. Finally, the paper concludes with a discussion of implications, limitations, and directions for further research.


\section{Proposed Change-Point Latent Factor Model for Item Response Data} 

\subsection{Background}
Most existing statistical methods used in educational testing assume no change in response style among the respondents. We begin by discussing such modeling approach. 

\subsection{Baseline IRT Measurement Model}

Consider a binary matrix $Y$ with $N$ rows and $J$ columns. The entries $Y_{ij} \in \{0, 1\}$ are random variables representing the response of respondent $i$ to item $j$, for $i=1, \ldots, N$ and $j=1, \ldots, J$. The response $Y_{ij} = 0$ corresponds to an incorrect response and $Y_{ij} = 1$ to a correct response. Traditional IRT models impose a joint distribution of $Y$ by making the following assumptions: $(i)$ each respondent $i$, for $i=1, \ldots, N$, is represented by a latent variable $\theta_i$, $(ii)$ the distribution of the response vector $\mathbf{Y}_i = (Y_{i1}, \ldots, Y_{iJ})^\top$ depends only on $\theta_i$ for given items, and $(iii)$ $Y_{i1}, \ldots, Y_{iJ}$ are conditionally independent given $\theta_i$. The first two assumptions directly inform the joint distribution of the responses and the latent trait, which can be expressed as $p(\mathbf{Y}_i, \theta_i | \boldsymbol{\eta}) = p(\mathbf{Y}_i | \theta_i; \boldsymbol{\eta}_1) p(\theta_i | \boldsymbol{\eta}_2) $, where $\boldsymbol{\eta} = (\boldsymbol{\eta}_1^\top, \boldsymbol{\eta}_2^\top)^\top$ represent the model parameters. Treating the latent variables $\theta_i$, $i=1, \ldots, N$, as independent and identically distributed random variables with density function with respect to the Lebesgue measure, it is possible to specify the density function for the observed variables as
\begin{equation}
p(\mathbf{Y}_i | \boldsymbol{\eta} ) = \int p(\mathbf{Y}_i | \theta_i; \boldsymbol{\eta}_1) p(\theta_i | \boldsymbol{\eta}_2) d \theta_i.
\end{equation}

The model $p(\mathbf{Y}_i | \theta_i; \boldsymbol{\eta}_1)$ is referred to as the measurement model and is determined by the parameters $\boldsymbol{\eta}_1 \in \mathcal{H}_1$. The model $p(\theta_i | \boldsymbol{\eta}_2)$, often referred to as the structural model, is determined by the parameters $\boldsymbol{\eta}_2 \in \mathcal{H}_2$. The latent variable $\theta_i$ can be either scalar or vector-valued and represents respondent $i$'s latent trait(s) that the test is designed to measure. In our notation and analysis, we will treat $\theta_i$ as unidimensional, with its distribution assumed to be standard normal, as is common in the IRT literature. The joint parameter space for $\boldsymbol{\eta}$ is $\mathcal{H} = \mathcal{H}_1 \times \mathcal{H}_2$, i.e., $\boldsymbol{\eta}_1$ and $\boldsymbol{\eta}_2$ are distinct.  

The measurement model $p(\mathbf{Y}_i | \theta_i; \boldsymbol{\eta}_1)$ is a function of item-level parameters and a parametric form is commonly assumed. A popular parametrization of the measurement model is
\begin{equation}\label{eq:IRT}
P(Y_{ij} = 1 | \theta_i; \boldsymbol{\eta}_{1}) = f_j(d_j + a_j \theta_i),
\end{equation}
where $f_j : \mathbbm{R} \rightarrow (0,1)$ is a pre-specified and monotonically increasing function. In educational testing, the parameters $(d_1, \ldots, d_J)$ are interpreted as the easiness of each item, and $(a_1, \ldots, a_J)$ represent item discrimination, i.e., how well an item discriminates between respondents with high and low levels of $\theta$. Here, $\boldsymbol{\eta}_{1j} = (d_j, a_j)$ represents the item parameters for item $j$, and collectively, $\boldsymbol{\eta}_1 = (\boldsymbol{\eta}_{11}, \ldots, \boldsymbol{\eta}_{1J})$ encompasses all item parameters. The structural model parameters $\boldsymbol{\eta}_2$ pertain to the distribution of $\theta_i$, which is assumed to be standard normal as mentioned earlier.

Popular choices for $f_j$ include the logistic link,
$$
f(x) = \frac{\exp(x)}{1 + \exp(x)},
$$
and the normal ogive link,
$$
f(x) = \int_{- \infty}^x \phi(z) dz, 
$$
where $\phi$ is the standard normal density function. Other choices of link function, such as the complimentary log-log link and functions with manually set lower and/or upper asymptotes, can also be accommodated.

When $f$ takes the logistic form, the model
\begin{equation}
f_{j}(1) = \frac{\exp(d_j + a_j \theta_i)}{1 + \exp(d_j + a_j \theta_i)}.
\end{equation}
is a reparameterization of the 2-Parameter Logistic (2-PL) model \citep{birnbaum1968some}. If the slope parameters $a_j$, $j=1, \ldots, J$, are constrained to be constant and equal to 1, the resulting model is a reparameterization of the popular Rasch model \citep{rasch1960studies},
\begin{equation}\label{eq:Rasch_baseline}
f_{j}(1) = \frac{\exp(d_j + \theta_i)}{1 + \exp(d_j + \theta_i)}.
\end{equation}
%


\subsection{IRT Model with Respondent-Level Change-Points}

We now discuss an extension of the measurement model in \eqref{eq:IRT} that includes respondent-level change-points to account for time-pressure effects. To define this model, we introduce an additional respondent-level latent variable, $\tau_i \in \mathcal{T}$. The variable $\tau_i$ is a discrete random variable with probability mass function with respect to the counting measure. It indicates, for each respondent, the item that marks the end of normal response behavior and the start of rapid response behavior due to time-pressure. The variable $\tau$ thus represents individual-level change-points. The largest possible value of $\tau_i$ is $J$, meaning that respondent $i$ does not have a change-point. We denote by $c$ the smallest possible value of $\tau$, and thus the last item not affected by any response behavior change. This point is naturally unknown, which will be further discussed in Section 3.

The change effect is characterised as a shift, denoted by $\gamma_j < 0$ for $j=1, \ldots, J$, in the intercept of the linear predictor in \eqref{eq:IRT} for post-change items. The measurement model therefore depends on both $\theta$ and $\tau$. That is,
\begin{equation}\label{eq:IRT_cp}
P(Y_{ij} = 1 \mid \theta_i, \tau_i; \boldsymbol{\eta}_1, \gamma_j) = 
f_{j}(d_j + a_j \theta_i + \mathbbm{1}_{\{j > \tau_i\}} \gamma_j) 
\end{equation}

In \eqref{eq:IRT_cp}, $\mathbbm{1}_{\{j > \tau_i\}}$ is an indicator function that is equal to 1 if item $j$ is after the respondent's change-point $\tau_i$, and $\gamma_j$ represents the change effect for item $j$. The proposed model thus indicates when the change takes place for each respondent and thereafter adjusts the model's intercept $d_j$ with $\gamma_j$ from the next item forward.

The baseline model $f_j(d_j + a_j \theta_i)$ models the item responses pre-change, and the extended model $f_{j}(d_j + a_j \theta_i + \mathbbm{1}_{\{j > \tau_i\}} \gamma_j)$ models the item responses post-change. Given the monotonicity of $f_j$ and that $\gamma_j < 0$ for all $j=1, \ldots, J$, it follows that 
$$
f_{j}(d_j + a_j \theta_i) > f_{j}(d_j + a_j \theta_i + \mathbbm{1}_{\{j > \tau_i\}} \gamma_j).
$$ 

To illustrate the proposed model, we provide an example using the 2-PL model as the baseline model. 

\subsubsection*{Example: 2-PL baseline model}
Using the 2-PL model with logit link as the baseline, pre-change item responses are modelled by
\begin{equation}\label{eq:2-PL_baseline}
P(Y_{ij} = 1 \mid \theta_i; \boldsymbol{\eta}_1)  = \frac{\exp(d_j + a_j \theta_i)}{1 + \exp(d_j + a_j \theta_i)}.
\end{equation}
Post-change item responses are modelled by
\begin{equation}\label{eq:2-PL_cp}
P(Y_{ij} = 1 \mid \theta_i, \tau_i; \boldsymbol{\eta}_1, \gamma_j)  = \frac{\exp(d_j + a_j \theta_i + \mathbbm{1}_{\{j > \tau_i\}} \gamma_j)}{1 + \exp(d_j + a_j \theta_i + \mathbbm{1}_{\{j > \tau_i\}} \gamma_j)}.
\end{equation}

\begin{figure}
		\includegraphics[width=\textwidth]{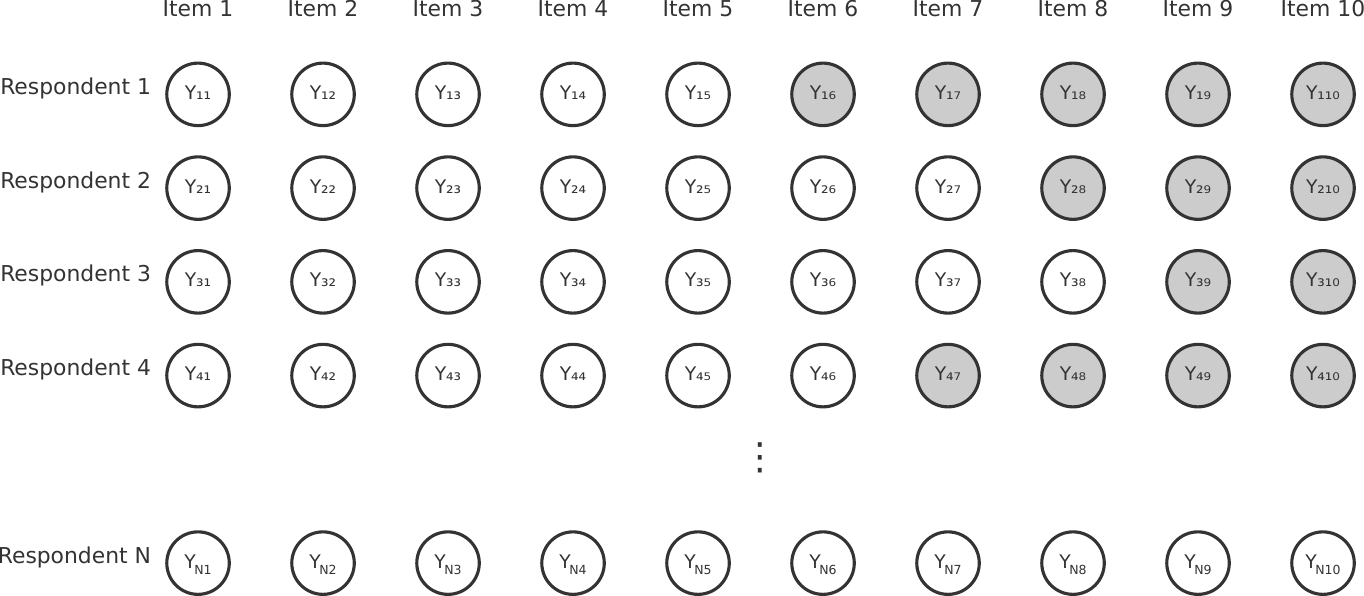}
		\caption{Visualization of the change-point model. The white circles represent normal item response behavior and the gray circles illustrate the change-points.}
		\label{fig:change-point-model}
	\end{figure}

In Figure \ref{fig:change-point-model}, we present an example illustrating how the change-points can be distributed across the items, for a 10-item test. For this example, the earliest change-point, $c$, is item 6, and Respondent 1 is the first test taker to change. Note that the change-points for the respondents are allowed to be located at different items, and that Respondent $N$ does not change at all in this particular example. For respondents with a change-point $c \leq \tau < J$, the post-change model contains $\gamma_j$-adjusted intercepts. For respondents such as Respondent $N$, the baseline model will characterise their whole response sequence.


 In Figure \ref{fig:path_diagram}, the proposed model is illustrated in a path diagram. The observed variables $Y_j$ are represented by squares and the latent variables $(\theta, \tau)$ by circles. For the first $c$ items, the item responses are only governed by the latent ability $\theta$. Post-change, the distribution of the responses are furthermore affected by $\tau$.

\begin{figure}[b!]
\begin{tikzpicture}[
node distance=1cm,
blue/.style={rectangle,draw=black,fill=white,minimum size=30pt},
black/.style={circle,draw=black,fill=white,minimum size=30pt},
red/.style={circle,draw=black,fill=black!10,minimum size=30pt}
]
\node[blue] (Y1) {$Y_1$};
\node[blue,right=of Y1] (Y2) {$Y_2$};
\node[blue,right=of Y2] (Ydots) {$\dots$};
\node[blue,right=of Ydots] (Yc) {$Y_c$};
\node[blue,right=of Yc] (Ycp1) {$Y_{c+1}$};
\node[blue,right=of Ycp1] (YJdots) {$\dots$};
\node[blue,right=of YJdots] (YJ) {$Y_J$};
\node[black,below right=1.8cm and 0.5cm of Y2] (theta) {$\theta$};
\node[red,below right=1.8cm and 0.25cm of Ycp1] (tau) {$\tau$};
\draw[->] (theta) -- (Y1);
\draw[->] (theta) -- (Y2);
\draw[->] (theta) -- (Ydots);
\draw[->] (theta) -- (Yc);
\draw[->] (theta) -- (Ycp1);
\draw[->] (theta) -- (YJdots);
\draw[->] (theta) -- (YJ);
\draw[->] (tau) -- (Ycp1);
\draw[->] (tau) -- (YJdots);
\draw[->] (tau) -- (YJ);
\end{tikzpicture}
\caption{Path diagram for the proposed model.}
\label{fig:path_diagram}
\end{figure}

\subsection{Structural Model}\label{sec:structural_model}
The structural model describes the joint distribution of $(\theta, \tau)$. We make the assumption that $\theta$ and $\tau$ are independent so that the joint distribution is given by the product of their marginal distributions. Given the continuous nature of $\theta$ and the discrete nature of $\tau$, their joint density function is therefore the Radon-Nikodym derivative with respect to the product measure composed of the Lebesgue measure and counting measure. In Section \ref{generalisations} we discuss model generalizations where e.g. the independence of $\theta$ and $\tau$ is relaxed. 

To define the marginal distribution of $\tau$, let the set of possible values of $\tau$ be defined by the ordered set \(\mathcal{T} = \{c, \ldots, J\}\). To characterize the distribution of the change-points, we employ a general logistic model:

$$
\log \left( \frac{P(\tau = j+1)}{P(\tau = j)} \right) = \alpha, \quad j \geq c,
$$
which implies:
$$
P(\tau = j) = p_c \cdot q^{j-c}, \quad j \geq c,
$$
where $q = e^{\alpha}$ and $p_c = P(\tau = c)$. Additionally, the probability of a respondent not having a change, i.e., the event $\tau = J$, is modeled as:
$$
\text{logit}(P(\tau=J)) = \beta \implies P(\tau=J) = p_J = \frac{e^{\beta}}{1 + e^{\beta}}.
$$
To ensure that these probabilities sum to 1, we normalize them with the factor:
$$
S = p_c \left( \frac{1 - q^{J-c}}{1 - q} \right) + p_J.
$$
Thus, the distribution of $\tau$ is given by:

\begin{equation}\label{eq:marg_tau}
p(\tau_i) = P(\tau_i = j) = 
\begin{cases}
\frac{p_c \cdot q^{j-c}}{S}, & \text{for } c \leq j < J, \\
\frac{p_J}{S}, & \text{for } j = J.
\end{cases}
\end{equation}

\begin{figure}
    \centering
    \includegraphics[width=\linewidth]{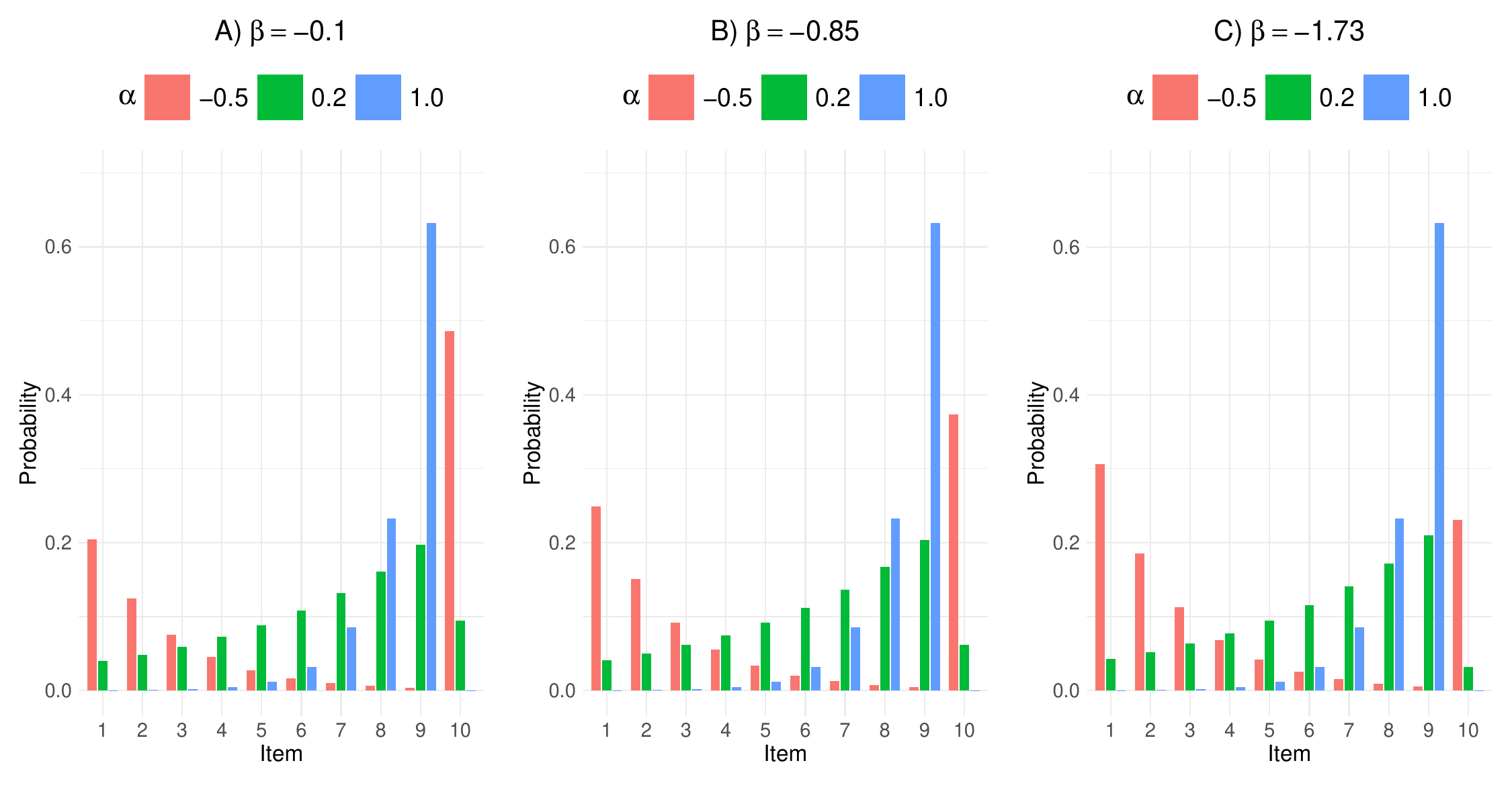}
    \caption{\textcolor{black}{Distribution of change-points $\tau$ for different values of $\alpha$ when $\beta$ is fixed at $-0.1$ (A), $-0.85$ (B), and $-1.73$ (C), respectively. The parameter $\alpha$ controls the relative likelihood of a change occurring at later versus earlier positions through a discrete-time hazard model, where $q = e^\alpha$ determines the ratio of probabilities between consecutive items. Higher values of $\alpha$ increase the likelihood that changes occur later in the sequence. The parameter $\beta$, which remains fixed in these plots, influences the overall probability of no change ($\tau = J$).}}
    \label{fig:change_point_distributions_alpha}
\end{figure}

\begin{figure}
    \centering
    \includegraphics[width=\linewidth]{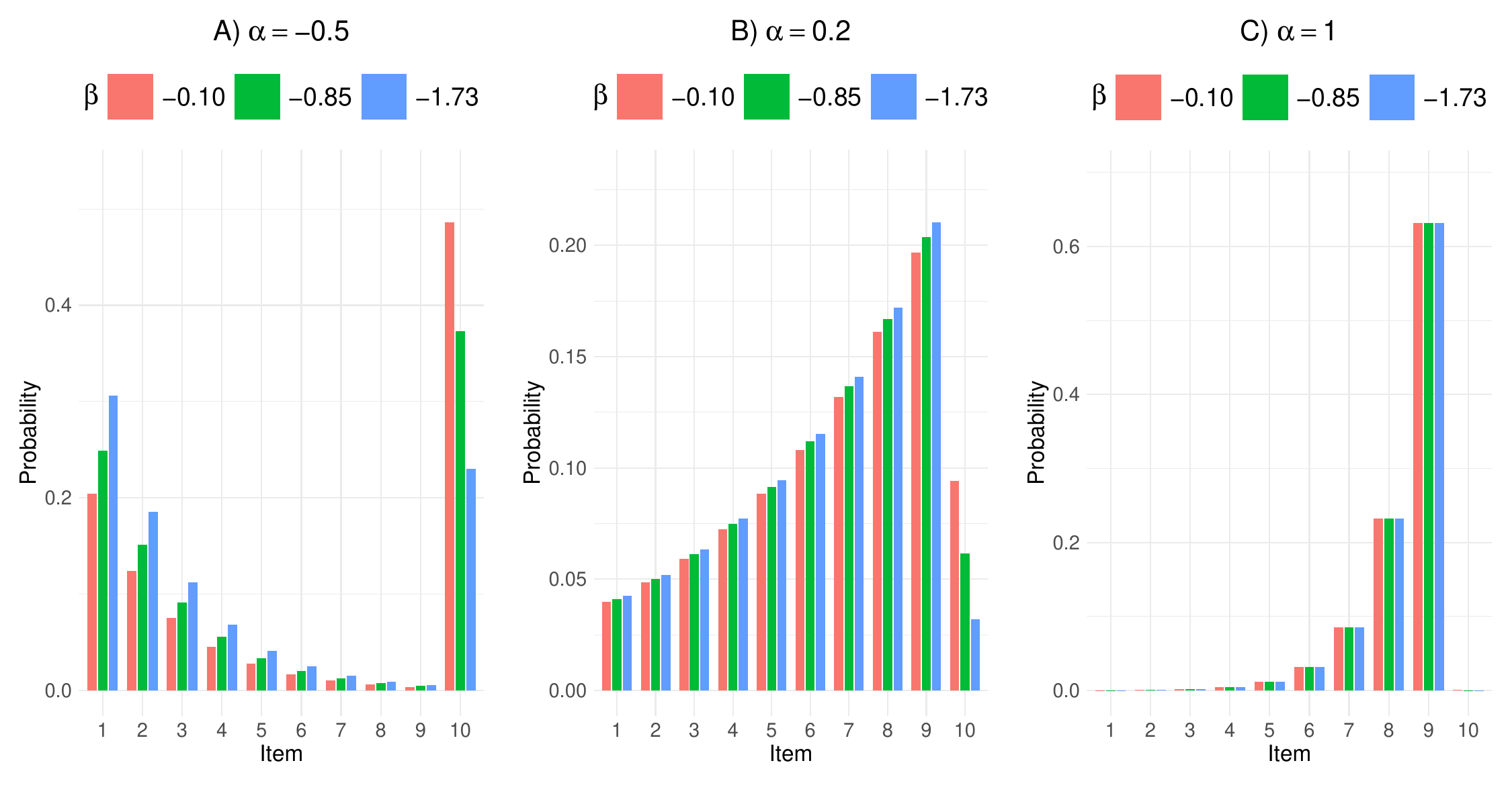}
    \caption{\textcolor{black}{Distribution of change-points $\tau$ for different values of $\beta$ when $\alpha$ is fixed at $-0.5$ (A), $0.2$ (B), and $1.0$ (C), respectively. The parameter $\beta$ controls the probability of no change ($\tau = J$) through a logistic function, with $P(\tau = J) = e^\beta/(1 + e^\beta)$. Higher values of $\beta$ increase the probability mass at the final time point $J$.}}
    \label{fig:change_point_distributions_beta}
\end{figure}

In this model, the change-points can be thought of as following a discrete-time hazard model, which is akin to a geometric distribution where the hazard (or probability of a change-point) is determined by the parameter $\alpha$. The parameter $\beta$ controls the probability that no change-point occurs, i.e., $\tau = J$, through a logistic regression model. A higher $\beta$ increases the probability that a respondent does not experience a change in response style. This distribution allows for a flexible approach that captures both the likelihood of change-points at various stages and the possibility of no change-point. We illustrate the change-point distribution for a few different values of $\alpha$ and $\beta$ in Figures \ref{fig:change_point_distributions_alpha} and \ref{fig:change_point_distributions_beta}. 

In this structural model, $\boldsymbol{\eta}_2$ encompasses the parameters that define the joint distribution of $(\theta, \tau)$. Specifically, $\boldsymbol{\eta}_2 = (\mu, \sigma^2, \alpha, \beta)$, where $\mu = 0$ and $\sigma^2 = 1$ define the standard normal distribution for $\theta$. The parameters $\mu$ and $\sigma^2$ are fixed to ensure model identifiability, while $\alpha$ and $\beta$ are estimated from the data.

	
\section{Statistical Inference}

\subsection{Marginal Likelihood}

The inference of the proposed model is based on the marginal likelihood function \citep{bock1981marginal}, where the item parameters $\boldsymbol{\eta}_1$ and the distribution of the person parameters $p(\theta, \tau; \boldsymbol{\eta}_2)$ are estimated simultaneously. Assuming that the responses \(Y_{ij}\) are conditionally independent across items \(j\) for each individual \(i\) given the latent trait \(\theta_i\) and change-point \(\tau_i\) (known as local independence), the marginal maximum likelihood estimator is given by
$$
(\hat{\boldsymbol{\eta}}_1, \hat{\boldsymbol{\eta}}_2) = \argmax_{\boldsymbol{\eta}_1, \boldsymbol{\eta}_2} \mathcal{L}(\boldsymbol{\eta}_1, \boldsymbol{\eta}_2) 
$$
where
\begin{align}\label{eq:llik_general}
\mathcal{L}(\boldsymbol{\eta}_1, \boldsymbol{\eta}_2) 
&= \prod_{i=1}^N \Bigg\{ \sum_{\tau \in \mathcal{T}} p(\tau \mid \boldsymbol{\eta}_2) \int \prod_{j=1}^J P(Y_{ij} = 1 \mid \theta, \tau; \boldsymbol{\eta}_1)^{y_{ij}} \nonumber \\
&\quad \times (1 - P(Y_{ij} = 1 \mid \theta, \tau; \boldsymbol{\eta}_1))^{1 - y_{ij}} \, p(\theta \mid \boldsymbol{\eta}_2) \, d\theta \Bigg\}.
\end{align}
The term $p(\tau_i \mid \boldsymbol{\eta}_2)$ denotes the probability mass function of the change-point $\tau_i$, and $p(\theta \mid \boldsymbol{\eta}_2)$ represents the density function of the latent trait $\theta_i$. The vector $\boldsymbol{\eta}_1$ includes the item-specific parameters while $\boldsymbol{\eta}_2$ comprises the structural parameters governing the distributions of $\theta$ and $\tau$. As explained in Section \ref{sec:structural_model}, a parametric form of both $\theta$ and $\tau$ is assumed. The estimator based on \eqref{eq:llik_general} can be viewed as an empirical Bayes estimator \citep{chen2021item}. 

With the 2-PL model as baseline measurement model, and when the marginal distribution of $\tau$ is given by equation \eqref{eq:marg_tau} and the marginal distribution of $\theta$ follows the standard normal distribution $\phi(\theta)$, the marginal likelihood function takes the following form:
\begin{equation}\label{eq:llik_specific}
\begin{split}
    \mathcal{L}(\boldsymbol{\eta}_1, \boldsymbol{\eta}_2) = \prod_{i=1}^N \Bigg\{ \sum_{\tau_i \in \mathcal{T}} p(\tau_i \mid \alpha, \beta) \int \left(\prod_{j=1}^J \frac{\exp\left(Y_{ij} (d_j + a_j \theta_i + \mathbbm{1}_{\{j > \tau_i\}} \gamma_j)\right)}{1 + \exp\left(d_j + a_j \theta_i + \mathbbm{1}_{\{j > \tau_i\}} \gamma_j\right)}\right) \phi(\theta) \, d\theta \Bigg\},
\end{split}
\end{equation}
where $\boldsymbol{\eta}_1 = (a_j, d_j, \gamma_j)$, $j=1, \ldots, J$, and $\boldsymbol{\eta}_2 = (\alpha, \beta)$.

Given the complexity of the marginal likelihood function, which involves both integration over the latent trait $\theta$ and summation over the change-point $\tau$, a closed-form analytical solution does not exist. Therefore, the maximization of the marginal likelihood function must be carried out using iterative numerical optimization methods. Various approaches can be employed for this purpose, including the Expectation-Maximization (EM) algorithm \citep{bock1981marginal, dempster1977maximum}, which is widely used in latent variable models. In our implementation, we utilize a quasi-Newton optimization method.

\subsection{Posterior Probability of Change-Points}

Given the model specification and the observed response data, we can compute the posterior probability of change-points for each respondent. For respondent $i$, the posterior probability of a change-point occurring at item $j$ is given by:
$$
P(\tau_i = j | \mathbf{Y}_i, \boldsymbol{\eta}_1, \boldsymbol{\eta}_2) = \frac{p(\mathbf{Y}_i | \tau_i = j, \boldsymbol{\eta}_1) p(\tau_i = j | \boldsymbol{\eta}_2)}{\sum_{k \in \mathcal{T}} p(\mathbf{Y}_i | \tau_i = k, \boldsymbol{\eta}_1) p(\tau_i = k | \boldsymbol{\eta}_2)}
$$
where $\mathbf{Y}_i = (Y_{i1}, \ldots, Y_{iJ})^\top$ is the response vector for respondent $i$, and $\mathcal{T}$ is the set of possible change-points. 
The likelihood $p(\mathbf{Y}_i | \tau_i = j, \boldsymbol{\eta}_1)$ can be computed by integrating over the latent trait $\theta_i$:
$$
p(\mathbf{Y}_i | \tau_i = j, \boldsymbol{\eta}_1) = \int p(\mathbf{Y}_i | \theta_i, \tau_i = j, \boldsymbol{\eta}_1) p(\theta_i|\boldsymbol{\eta}_2) d\theta_i.
$$
Using the 2-PL model with change-points as defined in equations \eqref{eq:2-PL_baseline} and \eqref{eq:2-PL_cp}, we can express this likelihood as:
$$
\begin{aligned}
p(\mathbf{Y}_i | \tau_i = j, \boldsymbol{\eta}_1) = \int &\prod_{k=1}^j \left(\frac{\exp(d_k + a_k \theta_i)}{1 + \exp(d_k + a_k \theta_i)}\right)^{Y_{ik}} \left(1 - \frac{\exp(d_k + a_k \theta_i)}{1 + \exp(d_k + a_k \theta_i)}\right)^{1-Y_{ik}} \\
&\times \prod_{k=j+1}^J \left(\frac{\exp(d_k + a_k \theta_i + \gamma_k)}{1 + \exp(d_k + a_k \theta_i + \gamma_k)}\right)^{Y_{ik}} \left(1 - \frac{\exp(d_k + a_k \theta_i + \gamma_k)}{1 + \exp(d_k + a_k \theta_i + \gamma_k)}\right)^{1-Y_{ik}} \\
&\times \phi(\theta_i) d\theta_i.
\end{aligned}
$$
The prior probability $p(\tau_i = j | \boldsymbol{\eta}_2)$ is given by equation \eqref{eq:marg_tau}. 

Computing these posterior probabilities allows us to assess the most likely position of the change-point for each respondent, as well as the uncertainty associated with this estimate. 

Furthermore, we can calculate the posterior probability that a change-point occurred at any point during the test for respondent $i$ as:
$$
P(\tau_i < J | \mathbf{Y}_i, \boldsymbol{\eta}_1, \boldsymbol{\eta}_2) = \sum_{j=1}^{J-1} P(\tau_i = j | \mathbf{Y}_i, \boldsymbol{\eta}_1, \boldsymbol{\eta}_2)
$$
This probability provides a summary measure of the evidence for a change in response style for each respondent, which can be used to identify respondents who may require further investigation or whose data may need special consideration in subsequent analyses.

\subsection{Model Selection}

For the proposed change-point IRT model, a key consideration is the determination of the earliest possible change-point, denoted by $c$. This parameter effectively constrains the model space and has implications for both model parsimony and interpretability. To this end, we employ a systematic approach:

\begin{enumerate}
\item Fit a sequence of models $\mathcal{M} = \{M_1, M_2, ..., M_K\}$, where each $M_k$ corresponds to a different value of $c$.

\item For each model $M_k$, compute an information criterion $IC(M_k)$. While various criteria exist, the Bayesian Information Criterion \citep[BIC;][]{schwarz1978estimating} is often preferred due to its consistency properties in model selection.

\item Select the optimal model $M^*$ that minimizes the chosen information criterion:

   $M^* = \argmin_{M_k \in \mathcal{M}} IC(M_k)$
\end{enumerate}

This approach allows for a data-driven selection of the most appropriate model within the specified family. 

It is worth noting that while this procedure focuses on selecting the optimal value of $c$, the same framework can be extended to compare models with different structural assumptions or parametrizations. For instance, one might consider different specifications of the change-point distribution $p(\tau)$. Furthermore, in the context of IRT models with change-points, it is important to consider not only statistical criteria but also substantive interpretability. The selected model should provide a meaningful understanding of the underlying response processes and potential shifts in respondent behavior throughout the test administration.

Sensitivity analyses can be conducted to assess the robustness of the model selection procedure. This may involve examining how the selected model changes under different information criteria or investigating the stability of parameter estimates across competing models. Ultimately, the model selection process serves to identify the most appropriate statistical representation of the change-points within the IRT framework, facilitating valid inferences about respondent behaviors and item characteristics in the presence of potential response style shifts.

\section{Model Generalizations}\label{generalisations}

The proposed change-point IRT model can be extended in several directions to accommodate more complex data structures and behavioral patterns. We outline several generalizations that enhance the model's flexibility and applicability.

\subsection{Latent Trait-Dependent Change-Points} 

One natural extension is to allow the change-point to depend on the latent trait $\theta_i$. This can be achieved by modifying the change-point distribution to incorporate $\theta_i$:

$$
P(\tau_i = j | \theta_i, \boldsymbol{\eta}_2) = 
\begin{cases}
\frac{p_c(\theta_i) \cdot q(\theta_i)^{j-c}}{S(\theta_i)}, & \text{for } j < J, \\
\frac{p_J(\theta_i)}{S(\theta_i)}, & \text{for } j = J,
\end{cases}
$$

where $p_c(\theta_i)$, $q(\theta_i)$, and $p_J(\theta_i)$ are now functions of $\theta_i$. For instance, we might specify:

$$
\begin{aligned}
\text{logit}(p_c(\theta_i)) &= \beta_0 + \beta_1\theta_i \\
\text{logit}(q(\theta_i)) &= \gamma_0 + \gamma_1\theta_i
\end{aligned}
$$

This formulation allows the probability of experiencing a change-point to vary with the respondent's latent trait level.

\subsection{Incorporation of Covariates} 

The model can be further extended to incorporate item-level or person-level covariates. For example, item response times could be integrated into the model. Let $t_{ij}$ denote the response time for person $i$ on item $j$. We can modify the item response function to include this information:

$$
P(Y_{ij} = 1 | \theta_i, \tau_i, t_{ij}; \boldsymbol{\eta}_1, \boldsymbol{\delta}) = \frac{\exp(d_j + a_j\theta_i + \mathbbm{1}_{\{j > \tau_i\}}\gamma_j + \delta_j\log(t_{ij}))}{1 + \exp(d_j + a_j\theta_i + \mathbbm{1}_{\{j > \tau_i\}}\gamma_j + \delta_j\log(t_{ij}))}
$$
where $\delta_j$ is a new item parameter capturing the effect of response time on the probability of a correct response.

\subsection{Extension to Polytomous Items} 

The model can be generalized to accommodate polytomous items using, for instance, the Graded Response Model \citep[GRM,][]{samejima1969estimation}. For an item $j$ with $K_j$ ordered categories, the GRM can be reparameterized to express the cumulative probability of responding in category $k$ or higher as:
$$
P(Y_{ij} \geq k | \theta_i, \tau_i; \boldsymbol{\eta}_1) = \frac{\exp(a_j\theta_i - b_{jk} + \mathbbm{1}_{\{j > \tau_i\}}\gamma_{jk})}{1 + \exp(a_j\theta_i - b_{jk} + \mathbbm{1}_{\{j > \tau_i\}}\gamma_{jk})}
$$
for $k = 1, \ldots, K_j - 1$, where $b_{jk}$ are category threshold parameters and $\gamma_{jk}$ are category-specific change effects.

\subsection{Multidimensional IRT Models} 

The change-point model can be extended to multidimensional IRT models. For a $D$-dimensional latent trait $\boldsymbol{\theta}_i = (\theta_{i1}, \ldots, \theta_{iD})$, we can specify:

$$
P(Y_{ij} = 1 | \boldsymbol{\theta}_i, \tau_i; \boldsymbol{\eta}_1) = \frac{\exp(d_j + \mathbf{a}_j^\top\boldsymbol{\theta}_i + \mathbbm{1}_{\{j > \tau_i\}}\boldsymbol{\gamma}_j^\top)}{1 + \exp(d_j + \mathbf{a}_j^\top\boldsymbol{\theta}_i + \mathbbm{1}_{\{j > \tau_i\}}\boldsymbol{\gamma}_j^\top)}
$$
where $\mathbf{a}_j = (a_{j1}, \ldots, a_{jD})$ and $\boldsymbol{\gamma}_j = (\gamma_{j1}, \ldots, \gamma_{jD})$ are vectors of discrimination and change effect parameters, respectively.

\subsection{Extension to 3-PL Model}

The model can be extended to incorporate a guessing parameter, akin to the 3-PL model:

$$
P(Y_{ij} = 1 | \theta_i, \tau_i; \boldsymbol{\eta}_1) = c_j + (1 - c_j)\frac{\exp(d_j + a_j\theta_i + \mathbbm{1}_{\{j > \tau_i\}}\gamma_j)}{1 + \exp(d_j + a_j\theta_i + \mathbbm{1}_{\{j > \tau_i\}}\gamma_j)}
$$
where $c_j$ is the guessing parameter for item $j$.

\subsection{Multiple Change-Points} 

The model can be generalized to allow for multiple change-points per respondent. Let $\boldsymbol{\tau}_i = (\tau_{i1}, \ldots, \tau_{iM})$ be a vector of $M$ change-points for respondent $i$. The item response function becomes:

$$
P(Y_{ij} = 1 | \theta_i, \boldsymbol{\tau}_i; \boldsymbol{\eta}_1) = \frac{\exp(d_j + a_j\theta_i + \sum_{m=1}^M \mathbbm{1}_{\{j > \tau_{im}\}}\gamma_{jm})}{1 + \exp(d_j + a_j\theta_i + \sum_{m=1}^M \mathbbm{1}_{\{j > \tau_{im}\}}\gamma_{jm})}
$$

where $\gamma_{jm}$ is the effect of the $m$-th change-point on item $j$.

These generalizations significantly expand the flexibility of the change-point IRT model, allowing it to capture a wide range of response behaviors and data structures. The choice of which extensions to incorporate will depend on the specific research questions and the nature of the data at hand.

\section{Simulation Study}

In this section, we describe the design and implementation of a simulation study to evaluate the performance of the proposed change-point latent factor model for item response data. We consider two simulation scenarios to assess parameter recovery and change-point estimation accuracy under different conditions. The first scenario treats the baseline parameters as known and focuses on the bias reduction in the $\theta$ estimates as we identify and adjust for the change-points. The second scenario treats all parameters as unknown and presents results for the parameter recovery when they are all estimated simultaneously. For all of our simulations, we consider the change-point 2-PL model in \eqref{eq:2-PL_baseline} and \eqref{eq:2-PL_cp}. 

\subsection{Simulation Design}

Our simulation study is designed to evaluate the performance of the proposed change-point latent factor model under controlled conditions. We generate data for a sample of 1,000 respondents answering 30 items. To investigate the impact of change-point locations, we consider two values for the earliest possible change-point, $c$, set at 20 and 25. This design allows us to examine how the model performs when changes in response behavior occur at different stages of the test.
The item parameters are generated to reflect realistic variations in item characteristics. The easiness parameters, $d_j$, are drawn from a Uniform(-1, 1) distribution, while the discrimination parameters, $a_j$, are sampled from a $\text{Uniform}(0.5, 1.5)$ distribution. These ranges are chosen to represent a diverse set of items with varying difficulty and discriminating power. The change effect parameters, $\gamma_j$, are set to 0 for items up to and including the change-point $c$, and are drawn from a $\text{Uniform}(-2, -1)$ distribution for subsequent items. This specification allows us to model a substantial shift in item difficulty post-change-point, reflecting the potential impact of time pressure on test performance.

To model the distribution of change-points across respondents, we employ the proposed discrete hazard model. The log-odds parameter $\alpha$ is fixed at 0.2, governing the spread of change-points across items. The logit parameter $\beta$ for the probability of no change-point is varied across three levels: -1.73, -0.85, and -0.1. These values correspond to approximate probabilities of no change-point of 15\%, 30\%, and 47.5\% \textcolor{black}{(derived directly from the logistic transformation exp($\beta$)/(1+exp($\beta$)))}, allowing us to examine the model's performance under different proportions of affected respondents.

\subsection{Simulation Scenarios}
We investigate two primary scenarios to comprehensively assess the model's capabilities. In the first scenario, we assume that the baseline item parameters ($d_j$ and $a_j$) are known, focusing on the $\theta$ estimates as we identify and adjust for the change-points. This scenario mimics situations in operational testing where item parameters have been pre-calibrated through extensive pretesting without time pressure. It allows us to isolate and evaluate the performance of the change-point detection mechanism when item characteristics are well-established.
The second scenario presents a more challenging and realistic setting where all parameters – item characteristics ($d_j$ and $a_j$), change effects ($\gamma_j$), and change-point distribution parameters ($\alpha$ and $\beta$) – are simultaneously estimated from the data. This task mirrors the complexities encountered in practical applications where prior item calibration may not be available.

By systematically varying the earliest possible change-point ($c$) and the proportion of respondents experiencing a change-point (through $\beta$), we aim to provide a thorough understanding of the model's behavior across a range of plausible testing situations. This simulation design enables us to evaluate the model's efficacy in recovering true parameters, its sensitivity to different change-point patterns, and its overall robustness in capturing the dynamics of response behavior changes in educational testing contexts.

\subsection{Evaluation Metrics}

We evaluate the performance of the item parameter estimates, change-point parameter estimates, and the latent trait \(\theta\) estimates using bias and root mean squared error (RMSE) as primary metrics. Additionally, we assess the accuracy of the change-point estimates \(\tau\) through the mean absolute error (MAE) between the estimated and true change-points across all respondents and simulation replications.

The MAE for the change-point estimates in a given simulation replication \(r\) is calculated as:
\[
\text{MAE}^{(r)}(\hat{\tau}) = \frac{1}{N} \sum_{i=1}^N |\hat{\tau}_i^{(r)} - \tau_i^{(r)}|,
\]
where \(\hat{\tau}_i\) and \(\tau_i\) are the estimated and true change-points for respondent \(i\), respectively.

The bias and RMSE for the latent trait \(\theta\) estimates across all respondents and simulation replications are defined as follows:
\[
\text{bias}(\hat{\theta}) = \frac{1}{N \times R} \sum_{r=1}^{R} \sum_{i=1}^{N} (\hat{\theta}_{ir} - \theta_{ir}),
\]
\[
\text{RMSE}(\hat{\theta}) = \frac{1}{R} \sum_{r=1}^{R} \sqrt{\frac{1}{N} \sum_{i=1}^{N} (\hat{\theta}_{ir} - \theta_{ir})^2},
\]
where \(R\) denotes the number of simulation replications, \(N\) is the number of respondents, \(\hat{\theta}_{ir}\) is the estimated \(\theta\) for respondent \(i\) in simulation replication \(r\), and \(\theta_{ir}\) is the corresponding true \(\theta\).

In addition to the standard $\theta$ estimates, we also compute adjusted $\theta$ estimates that are based only on a subset of items - specifically, those items up to the identified change-point for each respondent. This adjustment is motivated by the need to account for the possibility that responses after the change-point might be influenced by a different latent trait process, thereby potentially biasing the $\theta$ estimates if all items were used. By focusing only on the items up to the change-point, the adjusted estimates aim to more accurately reflect the respondent's latent trait before any potential shift.

For each respondent $i$ and simulation replication $r$, we calculate $\hat{\theta}_{ir}^{\text{subset}}$ using only the responses $Y_{ij}$ where $1 \leq j \leq \tau_{ir}$, where $\tau_{ir}$ is the estimated change-point for respondent $i$ in replication $r$. The bias and RMSE for these adjusted $\theta$ estimates are then defined as:
\[
\text{bias}(\hat{\theta}^{\text{subset}}) = \frac{1}{N \times R} \sum_{r=1}^{R} \sum_{i=1}^{N} (\hat{\theta}_{ir}^{\text{subset}} - \theta_{ir}),
\]

\[
\text{RMSE}(\hat{\theta}^{\text{subset}}) = \sqrt{\frac{1}{N \times R} \sum_{r=1}^{R} \sum_{i=1}^{N} (\hat{\theta}_{ir}^{\text{subset}} - \theta_{ir})^2}.
\]

Here, $\hat{\theta}_{ir}^{\text{subset}}$ represents the adjusted $\theta$ estimate for respondent $i$ in simulation replication $r$, calculated using only the responses up to the estimated change-point $\tau_{ir}$. The true $\theta_{ir}$ values are used as the reference for computing bias and RMSE.

For respondents with a change-point ($\tau_{ir} < J$), the bias and RMSE for the $\theta$ estimates are computed as:
\[
\text{bias}(\hat{\theta}^{\text{subset}}_{\text{cp}}) = \frac{1}{\sum_{r=1}^{R} \sum_{i=1}^{N} \mathbb{I}(\tau_{ir} < J)} \sum_{r=1}^{R} \sum_{i=1}^{N} \mathbb{I}(\tau_{ir} < J) (\hat{\theta}_{ir}^{\text{subset}} - \theta_{ir}),
\]

\[
\text{RMSE}(\hat{\theta}^{\text{subset}}_{\text{cp}}) = \sqrt{\frac{1}{\sum_{r=1}^{R} \sum_{i=1}^{N} \mathbb{I}(\tau_{ir} < J)} \sum_{r=1}^{R} \sum_{i=1}^{N} \mathbb{I}(\tau_{ir} < J) (\hat{\theta}_{ir}^{\text{subset}} - \theta_{ir})^2},
\]
where $\hat{\theta}_{ir}^{\text{subset}}$ represents the $\theta$ estimate for respondent $i$ in simulation replication $r$, calculated using only the responses up to the estimated change-point $\tau_{ir}$. The indicator function $\mathbb{I}(\tau_{ir} < J)$ ensures that only respondents with a change-point are included in these calculations.

Finally, the bias and RMSE for the item parameter estimates \( \hat{\zeta}_{j} \) (where \( \zeta_j \) represents the item parameters \( d_j \), \( a_j \), or \( \gamma_j \)) across the simulations are calculated as follows:
\[
\text{bias}(\hat{\zeta}_{j}) = \frac{1}{R} \sum_{r=1}^{R} \left( \hat{\zeta}_{jr} - \zeta_{j} \right),
\]
\[
\text{RMSE}(\hat{\zeta}_{j}) = \sqrt{\frac{1}{R} \sum_{r=1}^{R} \left( \hat{\zeta}_{jr} - \zeta_{j} \right)^2},
\]
where \( \hat{\zeta}_{jr} \) is the estimate of the item parameter \( \zeta_j \) in the \(r\)-th simulation replication, \( \zeta_{j} \) is the true parameter value, and \( R \) is the total number of simulation replications.

For all simulation settings, we treat the value of $c$ as known.

\subsection{Results}

\subsubsection{Scenario 1: Baseline Parameters Known}

In this scenario, we evaluated the bias and RMSE for the estimated latent variable, $\theta$, under different conditions, particularly focusing on various proportions of respondents with a change-point less than \( J \). We considered three levels of change-point probabilities: $\beta = -0.1$ (probability of no change-point approximately 0.475), $\beta = -0.85$ (probability approximately 0.30), and $\beta = -1.73$ (probability approximately 0.15). The results were calculated for both the full set of items and a subset where items affected by change-points were removed (referred to as ``after cleansing''). This approach enables an assessment of the impact of change-points on the accuracy of $\theta$ estimates and highlights the effectiveness of the cleansing procedure in mitigating these effects.

Table \ref{tab:theta_est_c20} presents the bias and RMSE for $\theta$ estimates across different $\beta$ values for $c=20$, both before and after cleansing. The results show that failing to account for change-points introduces a non-negligible bias in the $\theta$ estimates. Specifically, before cleansing, the bias for all respondents remains positive across all $\beta$ values, indicating a consistent overestimation of $\theta$. For instance, at $\beta = -0.1$, the bias is 0.118, and it increases to 0.196 at $\beta = -1.73$. This positive bias demonstrates the ability estimates are distorted if the change-points are not taken into account. The RMSE values before cleansing are also relatively higher, ranging from 0.383 to 0.397, further indicating that the presence of change-points contributes to increased estimation error. 

\begin{table}[]
\caption{The bias and RMSE for theta estimates, based on all items (Before cleansing) and items not affected by change-points (after cleansing) for various $\beta$ values and $c=20$. A $\beta$ value equal to -0.1 corresponds to 47.5\% of respondents with a change-point $<J$, $\beta=-0.85$ corresponds to approximately 30\% respondents with a change-point, and $\beta=-1.73$ corresponds to approximately 15\% respondents with a change-point.}
\begin{tabular}{@{}lcccccccc@{}}
\toprule
\multicolumn{1}{c}{$c=20$}             & \multicolumn{4}{c}{All respondents}                                        & \multicolumn{4}{c}{Speeded respondents}                                    \\ \midrule
\multicolumn{1}{c}{}             & \multicolumn{2}{c}{Before cleansing} & \multicolumn{2}{c}{After cleansing} & \multicolumn{2}{c}{Before cleansing} & \multicolumn{2}{c}{After cleansing} \\ \midrule
\multicolumn{1}{c}{CP parameter} & Bias              & RMSE             & Bias             & RMSE             & Bias              & RMSE             & Bias              & RMSE            \\ \midrule
$\beta = -0.1$ (47.5\%)             &   0.118           &    0.383         &    0.013         &   0.332          &        0.260      &       0.425      &        0.021      &         0.432   \\
$\beta=-0.85$ (approx. 30\%)        &    0.146          &     0.373        &    0.015         &      0.328       &         0.249     &     0.422        &      -0.017       &   0.420         \\
$\beta=-1.73$ (approx. 15\%)        &      0.196        &     0.397        &      -0.036       &     0.337        &        0.208      &      0.397       &    -0.025         &     0.406       \\ \bottomrule
\end{tabular}
\label{tab:theta_est_c20}
\end{table}

However, after the cleansing process, both bias and RMSE are substantially reduced. The bias after cleansing drops to near zero. For example, the bias at $\beta = -1.73$ shifts from 0.196 before cleansing to -0.036 after cleansing. Similarly, the RMSE consistently decreases after cleansing, with values dropping from 0.383 to 0.332 at $\beta = -0.1$ and from 0.397 to 0.337 at $\beta = -1.73$. This improvement in RMSE reflects the enhanced precision of the $\theta$ estimates when change-points are properly accounted for.

Focusing on only the respondents with a change, the results follow a similar trend, albeit with larger biases before cleansing. The bias for speeded respondents starts at 0.260 for $\beta = -0.1$ and remains substantial at 0.208 for $\beta = -1.73$. After cleansing, the bias for speeded respondents reduces significantly. This reduction demonstrates that the cleansing process is particularly important for improving the accuracy of $\theta$ estimates for speeded respondents, where the initial bias is more pronounced. The RMSE for changed respondents also decreases post-cleansing, although the reduction is less distinct compared to all respondents. This result suggests that while cleansing reduces estimation error, the complexity of speeded behavior still presents challenges for precise $\theta$ estimation.

\textcolor{black}{One aspect that may appear counterintuitive is the observed increase in bias before cleansing as the proportion of speeded respondents decreases. This trend can be explained by considering how the proportion of speeded respondents affects the estimation procedure. When there are more respondents with change-points (higher $\beta$), the model has more information to accurately estimate the change-point-related parameters, which in turn improves the overall estimation of $\theta$ for all respondents. However, as the proportion of speeded respondents decreases (lower $\beta$), the estimation of these parameters becomes more challenging due to the reduced number of informative cases. This can lead to increased estimation errors for all respondents, even those who do not exhibit a change in response behavior. Importantly, when focusing only on speeded respondents, the bias follows the expected trend - decreasing as the proportion of speeded respondents decreases.}

The results for $c=25$, presented in Table \ref{tab:theta_est_c25}, further illustrate the impact of accounting for change-points. Before cleansing, the bias for all respondents remains positive, though slightly lower than the biases observed at $c=20$. For example, the bias is 0.059 at $\beta = -0.1$ and increases to 0.145 at $\beta = -1.73$. These positive biases again highlight the systematic overestimation of $\theta$ when change-points are ignored. The RMSE values before cleansing, ranging from 0.357 to 0.369, also reflect this. After cleansing, there is a clear reduction in bias across all $\beta$ values. For instance, the bias at $\beta = -1.73$ decreases from 0.145 before cleansing to -0.023 after cleansing. The reduction in RMSE is also consistent, with the RMSE dropping from 0.357 to 0.316 at $\beta = -0.1$ and from 0.369 to 0.322 at $\beta = -1.73$. 

For respondents with a change-point, the biases before cleansing are again higher compared to all respondents, with a bias of 0.189 at $\beta = -0.1$ and 0.184 at $\beta = -1.73$. However, after cleansing, the bias for changed respondents is significantly reduced. The RMSE for speeded respondents also shows a reduction, though the decrease is slightly less pronounced compared to the results for all respondents.

The results from Tables \ref{tab:theta_est_c20} and \ref{tab:theta_est_c25} provide evidence that accounting for change-points is important for accurate $\theta$ estimation. Ignoring change-points introduces a substantial bias, leading to a systematic overestimation of ability levels. This bias is evident across all conditions before cleansing, and it is particularly pronounced for speeded respondents. The cleansing process, which removes items affected by change-points, mitigates this bias, often reducing it to near zero. The consistent reduction in RMSE across all conditions further demonstrates the enhanced precision of $\theta$ estimates when change-points are accounted for. These findings highlight the usefulness of incorporating change-point detection in IRT models, particularly in scenarios involving time constraints or other factors that might cause respondents to exhibit a change in behavior during the test.

In the next section, we will explore Scenario 2, where the baseline parameters $d_j$ and $a_j$ are estimated alongside the change-point and ability parameters.

\begin{table}[]
\caption{The bias and RMSE for theta estimates, based on all items (Before cleansing) and items not affected by change-points (after cleansing) for various $\beta$ values and $c=25$. A $\beta$ value equal to -0.1 corresponds to 47.5\% of respondents with a change-point $<J$, $\beta=-0.85$ corresponds to approximately 30\% respondents with a change-point, and $\beta=-1.73$ corresponds to approximately 15\% respondents with a change-point.}
\begin{tabular}{lcccccccc}
\hline
\multicolumn{1}{c}{$c=25$}         & \multicolumn{4}{c}{All respondents}                                        & \multicolumn{4}{c}{Speeded respondents}                                    \\ \hline
\multicolumn{1}{c}{}             & \multicolumn{2}{c}{Before cleansing} & \multicolumn{2}{c}{After cleansing} & \multicolumn{2}{c}{Before cleansing} & \multicolumn{2}{c}{After cleansing} \\ \hline
\multicolumn{1}{c}{CP parameter} & Bias              & RMSE             & Bias              & RMSE            & Bias              & RMSE             & Bias              & RMSE            \\ \hline
$\beta = -0.1$ (47.5\%)             &      0.059        &      0.357      &      0.010       &    0.316       &           0.189   &  0.388           &  0.029            &  0.413          \\
$\beta=-0.85$ (approx. 30\%)        &      0.112        &     0.364       &     -0.021        &       0.321    &           0.200   &    0.391         &    -0.037         &    0.395        \\
$\beta=-1.73$ (approx. 15\%)        &     0.145         &    0.369         &     -0.023        &      0.322      &        0.184      &    0.398        &   -0.039         &     0.401       \\ \hline
\end{tabular}
\label{tab:theta_est_c25}
\end{table}

\subsubsection{Scenario 2: All Parameters Unknown}
We focus here on the impact of varying the earliest possible change-point $c$ while keeping the change-point probability parameter $\beta$ fixed at -0.1, corresponding to approximately 47.5\% of respondents experiencing a change-point. Three values of $c$ were considered: 15, 20, and 25, with $J=30$ staying fixed. Additional scenarios exploring the effect of varying $\beta$ are presented in the Appendix. 

Tables \ref{tab:IRT_params_c15}, \ref{tab:IRT_params_c20}, and \ref{tab:IRT_params_c25} present the bias and RMSE of the IRT parameter estimates ($d$, $a$, and $\gamma$) for each item under the three $c$ values. The tables show that the recovery of the $d$ and $a$ parameters is generally good across all scenarios, with bias typically below 0.1 in absolute value and RMSE generally below 0.1. The estimation of the $\gamma$ parameters, which represent the change in item characteristics post-change-point, shows promising results. It is important to note that $\gamma$ parameters are only estimated for items after the change-point (i.e., for items $j > c$), as $c$ is treated as known in this simulation study. For these items, the bias in $\gamma$ estimates is generally low, often below 0.1 in absolute value, with RMSE typically ranging from 0.05 to 0.15. This indicates good recovery of the change effects, especially considering the complexity of estimating these additional parameters.

Comparing across different $c$ values, we observe that the estimation accuracy for $d$ and $a$ parameters remains relatively stable. However, for $\gamma$ parameters, there is a slight trend of increasing RMSE as $c$ increases, which is expected given the reduced number of items available for estimating the change effects. Despite this, even at $c=25$, the model still provides reasonable estimates of the $\gamma$ parameters, demonstrating its robustness across different change-point locations.

Table \ref{tab:combined_CP_params} presents the bias and root mean squared error (RMSE) of the change-point parameter estimates ($\alpha$ and $\beta$) for different values of $c$, along with the mean absolute error (MAE) for the estimated change-points ($\tau$). The $\alpha$ parameter, which controls the rate at which the probability of a change-point decreases across items, exhibits a small but increasing bias and RMSE as $c$ increases. The bias shifts from -0.008 (RMSE = 0.0179) at $c = 15$ to -0.071 (RMSE = 0.1033) at $c = 25$. This pattern suggests that when change-points are allowed to occur later in the test (larger $c$), there is less post-change information available to accurately estimate $\alpha$, leading to greater estimation uncertainty.

The $\beta$ parameter, which governs the probability of a respondent reaching the final item without a change-point, shows a small but positive bias across all scenarios, suggesting a tendency to slightly overestimate the likelihood of reaching the end of the test without a change. The bias in $\beta$ increases from 0.012 (RMSE = 0.0466) at $c = 15$ to 0.087 (RMSE = 0.1051) at $c = 20$, before slightly decreasing to 0.075 (RMSE = 0.1126) at $c = 25$. The increasing RMSE with $c$ indicates greater estimation variability as later change-points provide less post-change information.

The MAE for the estimated change-points ($\tau$) decreases as $c$ increases, from 2.095 when $c = 15$ to 1.694 at $c = 20$, and 1.126 at $c = 25$. This trend suggests that the model's precision in locating change-points improves when they occur later in the test. The decreasing MAE aligns with the observed patterns in $\alpha$ and $\beta$: as change-points shift later, there is less data available post-change, increasing uncertainty in $\alpha$ and $\beta$, while simultaneously making it easier to precisely locate the change-points due to a reduced range of possible locations.

Even in the most challenging scenario ($c = 15$), the average error in locating change-points is approximately 2 items, which remains reasonable given a test length of 30 items. These findings suggest that the proposed model can effectively detect and account for change-points in item responses across different values of $c$, maintaining reliable estimation even when change-points occur earlier in the test.

\begin{table}[ht]
\centering
\caption{Bias and RMSE of IRT Parameter Estimates when $c=15$ and $\beta=-0.1$.}
\label{tab:IRT_params_c15}
\begin{tabular}{ccccccc}
\hline
Parameter: & \multicolumn{2}{c}{$d$} & \multicolumn{2}{c}{$a$} & \multicolumn{2}{c}{$\gamma$} \\
Item & Bias & RMSE & Bias & RMSE & Bias & RMSE \\
\hline
1 & 0.034 & 0.039 & 0.018 & 0.056 & * & * \\
2 & 0.035 & 0.042 & 0.040 & 0.047 & * & * \\
3 & 0.050 & 0.055 & 0.036 & 0.054 & * & * \\
4 & 0.006 & 0.025 & -0.009 & 0.027 & * & * \\
5 & 0.049 & 0.054 & 0.048 & 0.054 & * & * \\
6 & 0.031 & 0.037 & 0.029 & 0.040 & * & * \\
7 & 0.045 & 0.051 & 0.054 & 0.060 & * & * \\
8 & -0.030 & 0.039 & -0.039 & 0.052 & * & * \\
9 & 0.030 & 0.038 & 0.009 & 0.033 & * & * \\
10 & 0.054 & 0.063 & 0.056 & 0.066 & * & * \\
11 & 0.018 & 0.022 & 0.013 & 0.035 & * & * \\
12 & -0.031 & 0.034 & -0.026 & 0.031 & * & * \\
13 & 0.061 & 0.067 & 0.047 & 0.052 & * & * \\
14 & -0.053 & 0.059 & -0.058 & 0.062 & * & * \\
15 & -0.033 & 0.039 & -0.043 & 0.057 & * & * \\
16 & -0.016 & 0.027 & -0.016 & 0.033 & -0.039 & 0.126 \\
17 & 0.000 & 0.025 & -0.010 & 0.034 & -0.012 & 0.123 \\
18 & 0.063 & 0.076 & 0.055 & 0.071 & 0.085 & 0.157 \\
19 & -0.075 & 0.078 & -0.090 & 0.093 & -0.110 & 0.158 \\
20 & 0.013 & 0.024 & -0.007 & 0.027 & 0.007 & 0.128 \\
21 & -0.040 & 0.048 & -0.042 & 0.048 & -0.073 & 0.149 \\
22 & 0.036 & 0.045 & 0.031 & 0.045 & 0.011 & 0.052 \\
23 & 0.024 & 0.028 & 0.024 & 0.034 & 0.021 & 0.107 \\
24 & -0.042 & 0.048 & -0.048 & 0.057 & -0.079 & 0.146 \\
25 & -0.002 & 0.041 & -0.014 & 0.028 & -0.029 & 0.095 \\
26 & 0.006 & 0.020 & -0.012 & 0.047 & -0.038 & 0.065 \\
27 & 0.063 & 0.074 & 0.070 & 0.077 & 0.045 & 0.066 \\
28 & 0.049 & 0.054 & 0.045 & 0.058 & 0.026 & 0.035 \\
29 & 0.062 & 0.074 & 0.061 & 0.076 & 0.006 & 0.069 \\
30 & 0.041 & 0.049 & 0.036 & 0.053 & -0.023 & 0.080 \\
\hline
\end{tabular}
\end{table}

\begin{table}[ht]
\centering
\caption{Bias and RMSE of IRT Parameter Estimates when $c=20$ and $\beta=-0.1$.}
\label{tab:IRT_params_c20}
\begin{tabular}{ccccccc}
\hline
Parameter: & \multicolumn{2}{c}{$d$} & \multicolumn{2}{c}{$a$} & \multicolumn{2}{c}{$\gamma$} \\
Item & Bias & RMSE & Bias & RMSE & Bias & RMSE \\
\hline
1 & -0.006 & 0.027 & -0.003 & 0.037 & * & * \\
2 & 0.003 & 0.024 & 0.010 & 0.025 & * & * \\
3 & 0.072 & 0.082 & 0.074 & 0.087 & * & * \\
4 & 0.009 & 0.022 & -0.011 & 0.059 & * & * \\
5 & 0.043 & 0.054 & 0.053 & 0.065 & * & * \\
6 & 0.027 & 0.038 & 0.041 & 0.059 & * & * \\
7 & 0.006 & 0.025 & 0.010 & 0.029 & * & * \\
8 & 0.008 & 0.029 & 0.004 & 0.052 & * & * \\
9 & -0.038 & 0.045 & -0.062 & 0.071 & * & * \\
10 & -0.054 & 0.063 & -0.054 & 0.061 & * & * \\
11 & 0.013 & 0.026 & 0.014 & 0.042 & * & * \\
12 & 0.023 & 0.033 & 0.031 & 0.040 & * & * \\
13 & -0.007 & 0.031 & -0.020 & 0.057 & * & * \\
14 & 0.029 & 0.038 & 0.040 & 0.048 & * & * \\
15 & 0.028 & 0.044 & 0.006 & 0.047 & * & * \\
16 & -0.001 & 0.024 & 0.008 & 0.026 & * & * \\
17 & 0.019 & 0.030 & -0.005 & 0.054 & * & * \\
18 & 0.041 & 0.057 & 0.047 & 0.067 & * & * \\
19 & 0.022 & 0.034 & 0.004 & 0.060 & * & * \\
20 & 0.036 & 0.052 & 0.020 & 0.050 & * & * \\
21 & 0.027 & 0.037 & 0.028 & 0.042 & 0.028 & 0.145 \\
22 & 0.076 & 0.086 & 0.059 & 0.074 & 0.063 & 0.116 \\
23 & 0.023 & 0.037 & 0.016 & 0.032 & 0.008 & 0.141 \\
24 & -0.008 & 0.031 & -0.015 & 0.040 & -0.073 & 0.098 \\
25 & 0.004 & 0.022 & -0.025 & 0.073 & -0.020 & 0.129 \\
26 & 0.034 & 0.042 & 0.005 & 0.058 & -0.022 & 0.059 \\
27 & 0.013 & 0.032 & -0.005 & 0.049 & -0.046 & 0.076 \\
28 & 0.016 & 0.031 & 0.003 & 0.042 & -0.037 & 0.073 \\
29 & 0.045 & 0.056 & -0.002 & 0.055 & -0.078 & 0.139 \\
30 & -0.020 & 0.040 & -0.059 & 0.078 & -0.090 & 0.117 \\
\hline
\end{tabular}
\end{table}

\begin{table}[ht]
\centering
\caption{Bias and RMSE of IRT Parameter Estimates when $c=25$ and $\beta=-0.1$.}
\label{tab:IRT_params_c25}
\begin{tabular}{ccccccc}
\hline
Parameter: & \multicolumn{2}{c}{$d$} & \multicolumn{2}{c}{$a$} & \multicolumn{2}{c}{$\gamma$} \\
Item & Bias & RMSE & Bias & RMSE & Bias & RMSE \\
\hline
1 & 0.002 & 0.027 & -0.006 & 0.038 & * & * \\
2 & 0.015 & 0.027 & 0.020 & 0.041 & * & * \\
3 & 0.088 & 0.097 & 0.087 & 0.099 & * & * \\
4 & -0.053 & 0.061 & -0.096 & 0.106 & * & * \\
5 & 0.036 & 0.052 & 0.034 & 0.051 & * & * \\
6 & 0.033 & 0.046 & 0.023 & 0.049 & * & * \\
7 & 0.039 & 0.049 & 0.042 & 0.059 & * & * \\
8 & 0.037 & 0.056 & 0.014 & 0.046 & * & * \\
9 & 0.004 & 0.034 & -0.055 & 0.094 & * & * \\
10 & 0.043 & 0.054 & 0.042 & 0.060 & * & * \\
11 & 0.025 & 0.041 & 0.022 & 0.051 & * & * \\
12 & 0.046 & 0.055 & 0.046 & 0.064 & * & * \\
13 & -0.027 & 0.037 & -0.061 & 0.075 & * & * \\
14 & 0.033 & 0.039 & 0.034 & 0.048 & * & * \\
15 & 0.051 & 0.072 & 0.020 & 0.061 & * & * \\
16 & 0.018 & 0.031 & 0.023 & 0.039 & * & * \\
17 & -0.017 & 0.035 & -0.049 & 0.067 & * & * \\
18 & 0.044 & 0.061 & 0.043 & 0.067 & * & * \\
19 & -0.052 & 0.059 & -0.060 & 0.072 & * & * \\
20 & -0.021 & 0.039 & -0.059 & 0.069 & * & * \\
21 & -0.011 & 0.025 & -0.004 & 0.031 & * & * \\
22 & -0.003 & 0.026 & -0.026 & 0.047 & * & * \\
23 & 0.061 & 0.078 & 0.062 & 0.077 & * & * \\
24 & -0.054 & 0.061 & -0.065 & 0.075 & * & * \\
25 & 0.007 & 0.038 & -0.016 & 0.036 & * & * \\
26 & -0.039 & 0.047 & -0.059 & 0.074 & -0.093 & 0.108 \\
27 & 0.022 & 0.061 & 0.033 & 0.059 & -0.052 & 0.132 \\
28 & 0.038 & 0.072 & 0.019 & 0.056 & -0.044 & 0.112 \\
29 & -0.029 & 0.042 & -0.055 & 0.070 & -0.085 & 0.111 \\
30 & 0.006 & 0.043 & 0.000 & 0.046 & -0.087 & 0.149 \\
\hline
\end{tabular}
\end{table}

\begin{table}[ht]
\centering
\caption{Bias and RMSE of change-point parameter estimates for different values of $c$ and $\beta=-0.1$.}
\label{tab:combined_CP_params}
\begin{tabular}{|c|c|c|c|c|c|c|}
\hline
\multirow{2}{*}{\textbf{Parameter}} & \multicolumn{2}{|c|}{$c=15$} & \multicolumn{2}{|c|}{$c=20$} & \multicolumn{2}{|c|}{$c=25$} \\
\cline{2-7}
 & \textbf{Bias} & \textbf{RMSE} & \textbf{Bias} & \textbf{RMSE} & \textbf{Bias} & \textbf{RMSE} \\
\hline
$\alpha$ & -0.008 & 0.0179 & -0.047 & 0.0592 & -0.071 & 0.1033 \\
$\beta$  & 0.012  & 0.0466 & 0.087  & 0.1051 & 0.075  & 0.1126 \\
\hline
\textbf{MAE}($\tau$) & \multicolumn{2}{c|}{2.095} & \multicolumn{2}{c|}{1.694} & \multicolumn{2}{c|}{1.126} \\
\hline
\end{tabular}
\end{table}

\section{Applications to Educational Testing Data}
	
In this section, we apply the proposed model to two real educational testing datasets. The aim is to empirically validate the model's ability to detect change-points in response styles and to estimate the associated item and change-point parameters accurately. Both datasets are from educational tests constructed to measure quantitative skills. For the first dataset, we have access to the response time of each respondent to each item, which gives an opportunity to validate our findings to some extent. 
	
\subsection{Dataset 1}

Dataset 1 comes from a quantitative section of a high-stakes standardized test widely used in the United States and many other countries\footnote{Source: Derived from data provided by ETS. Copyright © 2024 ETS. www.ets.org. The opinions set forth in this publication are those of the authors and not ETS.}. Many test-takers invest considerable time and resources in preparation, as their scores can significantly influence their educational and career opportunities. However, it is worth noting that test-takers may approach it with varying levels of motivation and preparation. Some may view it as a critical part in their academic journey, while others might see it as just one component of their overall application.

The dataset contains responses from $N=2,568$ respondents to $J=28$ items. We started by setting up a grid of $c$-values: $c = \{ J-1, J-2, ..., J/2 \}$, and fitted the model for each value of $c$, including a model without any change-points. Based on the BIC, we selected $c=19$\footnote{The BIC  for the chosen model equals 71896.48. The BIC of a model without any change-points i.e., the normal 2-PL model equals 72026.49}. Therefore, the earliest change-point is at item 19, and the intercept parameter $d_j$ is adjusted by $\gamma_j$ from item 20 and onwards for respondents with a change. 
	
The left-hand side of Table \ref{tab:IRT_params_combined} presents the item parameter estimates. The estimated $d_j$ values range from $-0.967$ to $3.076$, indicating significant variation in item easiness. Higher estimated $d_j$ values suggest easier items, with item 9 being estimated the easiest ($\hat{d}_9 = 3.076$) and item 26 estimated the most difficult ($\hat{d}_{26} = -0.967$). The estimated $a_j$ values range from $0.456$ to $2.052$, indicating varying levels of item discrimination. For $j > 19$, the estimated $\gamma_j$ varies, with values ranging from $-0.791$ to $-3.349$. This suggests a notable shift in response style, with item 23 showing the most decrease in the log odds of a positive response after the change-point ($\hat{\gamma}_{23} = -3.349$). We believe the relatively large, negative $\hat{\gamma}_j$ values are a reflection of the large number of zero values for $Y_{ij}$ towards the end of the test, as a result of the time pressure.
	
\begin{table}
    \centering
    \caption{Parameter estimates for the measurement models in the two educational testing data applications.}
    \begin{tabular}{|c|c|c|c|c|c|c|c|}
        \hline
        Item & \multicolumn{3}{c|}{Dataset 1} & & \multicolumn{3}{c|}{Dataset 2} \\
        \hline
        & $d_j$ & $a_j$ & $\gamma_j$ & & $d_j$ & $a_j$ & $\gamma_j$ \\
        \hline
        1  &  2.118 &  1.422 &  * &  &  1.322 &  0.856 &  * \\
        2  &  0.768 &  1.371 &  * &  &  1.067 &  1.193 &  * \\
        3  &  1.912 &  0.930 &  * &  &  1.736 &  1.254 &  * \\
        4  &  2.532 &  1.219 &  * &  &  1.088 &  0.865 &  * \\
        5  &  1.791 &  1.663 &  * &  & -0.046 &  0.958 &  * \\
        6  &  2.367 &  1.630 &  * &  & -0.292 &  1.043 &  * \\
        7  &  0.820 &  1.525 &  * &  & -0.255 &  1.264 &  * \\
        8  &  2.213 &  1.755 &  * &  & -0.923 &  1.124 &  * \\
        9  &  3.076 &  1.605 &  * &  & -0.721 &  1.122 &  * \\
        10 &  1.401 &  1.693 &  * &  & -0.491 &  0.995 &  * \\
        11 &  1.995 &  1.346 &  * &  & -0.974 &  0.779 &  * \\
        12 &  0.137 &  0.971 &  * &  & -0.193 &  0.709 &  * \\
        13 &  2.012 &  0.985 &  * &  &  1.747 &  0.730 &  * \\
        14 &  2.195 &  1.387 &  * &  &  0.541 &  0.907 &  * \\
        15 &  1.423 &  1.799 &  * &  &  0.728 &  0.698 &  * \\
        16 & -0.709 &  0.698 &  * &  & -0.856 &  0.747 &  * \\
        17 &  0.166 &  1.556 &  * &  & -0.378 &  0.220 &  * \\
        18 & -0.903 &  0.990 &  * &  & -0.817 &  1.080 &  * \\
        19 &  0.896 &  1.581 &  * &  &  0.780 &  1.392 &  * \\
        20 &  0.305 &  1.914 & -3.144 &  & -0.626 &  1.205 &  * \\
        21 & -0.546 &  2.052 & -1.016 &  &  0.435 &  0.970 &  * \\
        22 &  0.402 &  1.023 & -0.869 &  &  0.769 &  0.977 &  * \\
        23 & -0.222 &  1.848 & -3.349 &  & -0.471 &  1.175 & -3.681 \\
        24 &  0.120 &  1.691 & -2.475 &  & -1.275 &  0.796 & -3.036 \\
        25 &  0.097 &  0.784 & -1.280 &  &  0.370 &  0.891 & -2.929 \\
        26 & -0.967 &  0.845 & -0.791 &  &  1.603 &  0.859 & -2.025 \\
        27 & -0.191 &  0.845 & -1.978 &  & -0.522 &  1.158 & -1.390 \\
        28 & -0.136 &  0.456 & -1.623 &  &  1.303 &  1.835 & -3.849 \\
        29 &        &        &        &  & -0.710 &  0.775 & -0.965 \\
        30 &        &        &        &  &  0.180 &  0.625 & -2.144 \\
        31 &        &        &        &  &  1.128 &  0.781 & -2.037 \\
        32 &        &        &        &  & -0.811 &  1.016 & -2.071 \\
        33 &        &        &        &  & -0.236 &  0.925 & -3.093 \\
        34 &        &        &        &  & -0.733 &  1.044 & -3.591 \\
        35 &        &        &        &  &  0.440 &  1.161 & -8.231 \\
        36 &        &        &        &  & -0.165 &  1.285 & -3.041 \\
        \hline
    \end{tabular}
    \label{tab:IRT_params_combined}
\end{table}
	
\begin{figure}
		\centering
		\begin{subfigure}[b]{0.45\textwidth}
			\includegraphics[width=\textwidth]{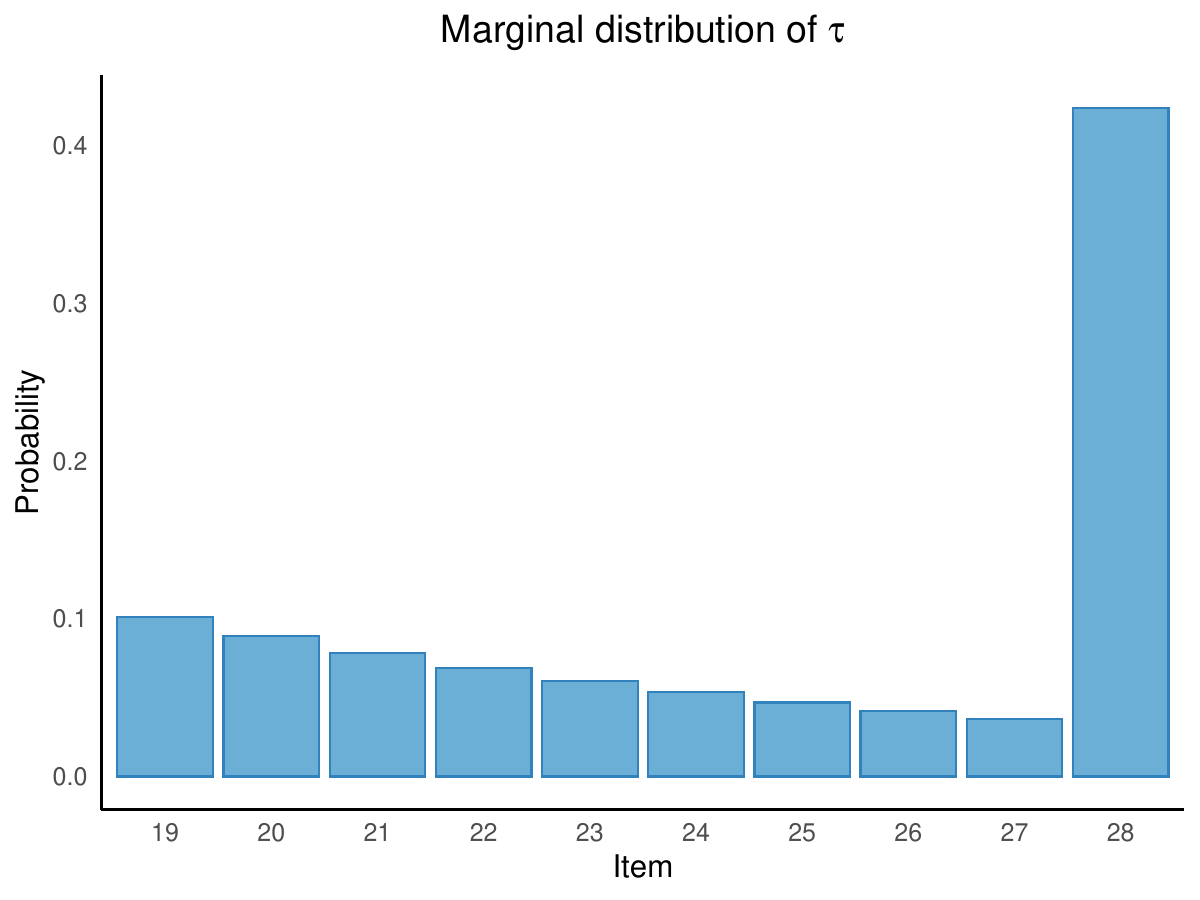}
			\caption{Estimated marginal distribution of $\tau$ for the first empirical dataset.}
			\label{fig:prior_probs}
		\end{subfigure}
		\hfill
		\begin{subfigure}[b]{0.45\textwidth}
			\includegraphics[width=\textwidth]{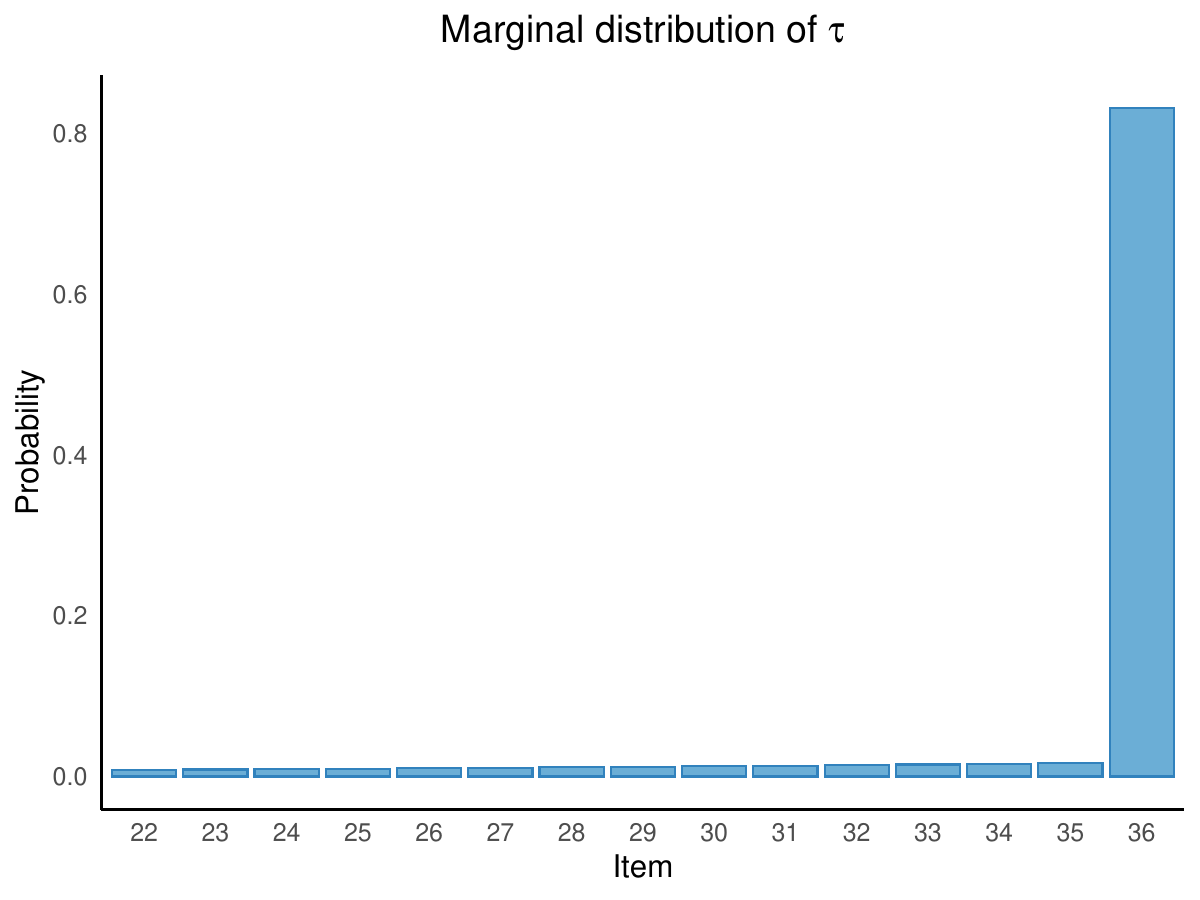}
			\caption{Estimated marginal distribution of $\tau$ for the second empirical dataset.}
			\label{fig:posterior_mode}
		\end{subfigure}
		\caption{Estimated marginal distribution of $\tau$ for the two empirical datasets.}
		\label{fig:prior_comparison}
	\end{figure}

The change-point parameter estimates are summarized in Table \ref{tab:change_point_params}. The estimated $\alpha$ of $-0.128$ indicates a slight tendency for change-points to occur earlier rather than later in the test. The estimated $\beta$ of $-0.307$ suggests a lower probability of respondents maintaining a consistent response style across all items, with a tendency towards adopting a new response style after item 19. In Figure \ref{fig:prior_comparison}, the estimated marginal distribution of $\tau$ (left-hand side) is displayed. \textcolor{black}{The pattern in Figure 5(a), where $\alpha < 0$, shows a peak in change-point probability at an early position followed by decreasing probabilities. This reflects a scenario where the hazard of experiencing a change-point decreases over time, suggesting that if respondents modify their response behavior, they are more likely to do so at a particular early point in the test rather than gradually over time. In contrast, Figure 5(b), where $\alpha > 0$, shows a slightly increasing hazard of change-points across items, indicating a scenario where respondents become more likely to change their response behavior as they progress through the test. In both cases, we observe the highest probability at the last item, representing respondents who did not experience a change-point during the test. The marginal distributions illustrated in Figure 5(a) and Figure 5(b) both align with the marginal distributions in Figure 3 and 4, where the proposed change-point model is illustrated for varying $\alpha$ and $\beta$ parameters.}

	\begin{table}
		\centering
		\caption{Estimated Change-Point Parameters}
		\begin{tabular}{|c|c|c|c|c|}
			\hline
			Dataset & $J$ & $c$ & $\hat{\alpha}$ & $\hat{\beta}$ \\
			\hline
			1 & 28 & 19 & -0.128 & -0.307 \\
			2 & 36 & 22 & 0.055 & 1.6 \\
			\hline
		\end{tabular}
		\label{tab:change_point_params}
	\end{table}

\begin{figure}
		\centering
		\includegraphics[scale=0.7]{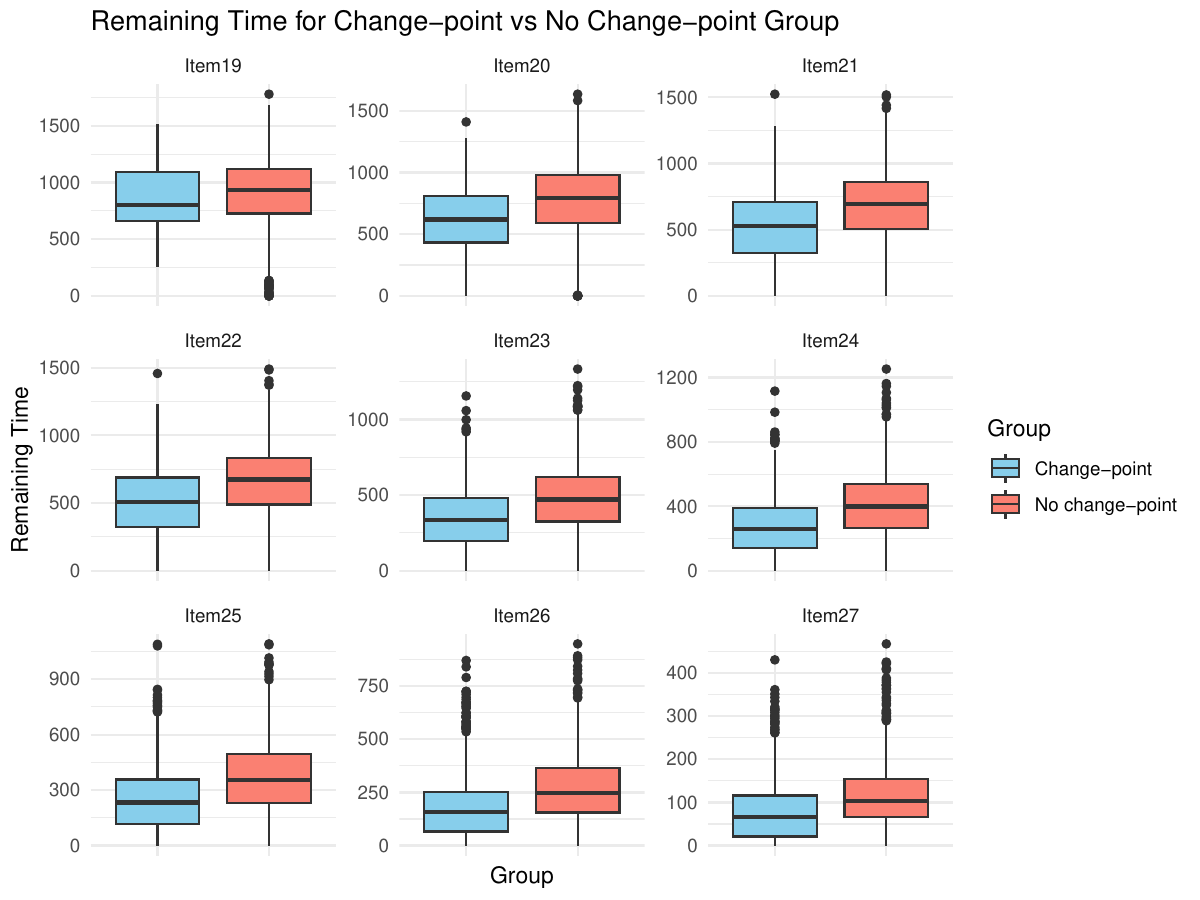}
		\caption{The remaining time for the change-point and no-change-point group.}
		\label{fig:remaining_time}
	\end{figure}

Figure \ref{fig:remaining_time} compares the remaining time of the test for respondents with and without a change-point across items 19 to 27. The boxplots for each item show the distribution of the remaining time after the completion of the specified item. Across all items, there is a noticeable variation in the remaining time between the ``Change-point'' and ``No change-point'' groups. This variation highlights the impact of encountering a change-point on the time management of respondents. It is also evident that the median remaining time for the ``Change-point'' group is lower than that of the ``No change-point'' group. This indicates that respondents who encounter a change-point may spend more time on these items, leaving them with less time for the remaining items. Indeed, there is a noticeable reduction in the remaining time for the ``Change-point'' group, especially for items 25 and 27. The ``No change-point'' group maintains a relatively stable median remaining time across these items.

\subsection{Dataset 2}

Dataset 2 comes from a mathematics placement test administered at a university in the United States. The test is designed to inform course selection decisions. For example, the results may determine whether a student begins with introductory calculus or advances directly to third-semester calculus. The stakes of this test can be considered moderate and vary among test-takers. For some students aiming to bypass introductory courses, the test may be perceived as high-stakes. Others may approach the test with less concern about their starting course level or may simply seek an honest assessment of their abilities. There is also a possibility that some students might intentionally underperform to be placed in easier courses. Generally, this test can be viewed as having lower stakes than the high-stakes test of Dataset 1, but higher stakes than state assessment tests, which typically have no direct consequences for students. The results of this placement test do have effects on a student's academic path, even if these effects are less substantial than those of traditional high-stakes tests.
	
The data consists of $N=3,000$ respondents answering $J=36$ items. Using the same BIC procedure as with the previous dataset, $c$ was selected to be 22. In the right-hand side of Table \ref{tab:IRT_params_combined}, the item parameters are displayed. The negative values of estimated \(\gamma_j\) suggests that item difficulty might be overestimated as respondents progress through the test and start experiencing time pressure. As was the case for the first dataset, we note that the magnitude of the $\gamma$ estimates are affected by the 0 pattern towards the end due to the time constraints.
	
The estimated change-point parameters $\alpha$ and $\beta$ are displayed in Table \ref{tab:change_point_params}. The positive value of $\hat{\alpha} = 0.055$ indicates that the log-odds of the change-point occurring at item $j+1$ relative to item $j$ slightly increases as we move further along the test. This implies that change-points are somewhat more likely to occur in later items, which differs from the pattern observed in Dataset 1. The estimate $\hat{\beta} = 1.6$ suggests a moderately high probability that respondents will not experience any change-points during the test. To interpret these parameters in the context of the model, recall that $q = e^{\alpha} \approx 1.057$, indicating a slight increase in the probability of a change-point occurring as the test progresses. The probability of no change-point ($p_J$) is given by the logistic function of $\beta$, which equals approximately 0.832. 

We illustrate the estimated marginal distribution of $\tau$ in Figure \ref{fig:prior_comparison}. This distribution shows a slightly increasing trend for the probability of change-points across items, with a substantial spike at $\tau = J$ representing the high probability of no change-point occurring. This analysis suggests that while change-points are slightly more likely to occur later in the test, the majority of respondents are still expected to complete the test without experiencing a change-point.

\section{Discussion}

This paper introduces a novel change-point latent factor model for item response data, specifically designed to address the challenge of detecting and accounting for shifts in response behavior during educational testing. Our model extends traditional IRT frameworks by incorporating individual-level change-points, allowing for a more nuanced understanding of how test-takers' response patterns may evolve over the course of an assessment. The primary contribution of this work lies in its ability to simultaneously estimate item parameters, a person latent trait, and change-point locations, providing a comprehensive approach to modeling the complex dynamics of test-taking behavior.

\textcolor{black}{A strength of our modeling approach lies in its explicit incorporation of a temporal structure through the discrete-time hazard model, in contrast to approaches such as in \cite{wang2015mixture} that only models the presence or absence of changes through a Bernoulli distribution. While such models can identify whether a change occurred, our framework provides richer insights by modeling when these changes are likely to occur. This temporal modeling is particularly valuable in educational testing contexts, where the timing of behavioral changes may carry important diagnostic information - for example, distinguishing between early changes that might reflect students' initial adjustment to the testing environment and later changes that could indicate speededness effects. The model's parameters ($\alpha$ and $\beta$) allow researchers and practitioners to better understand the dynamics of test-taking behavior. This approach enables not only the detection of behavioral changes but also insights into their temporal distribution, which can be valuable for test development and administration.}

Our simulation studies and empirical analyses of two real-world educational testing datasets demonstrate the efficacy and practical utility of the proposed model. Findings from these investigations highlight several important aspects of our approach. The simulation results show that our model can accurately recover item parameters, change-point parameters, and individual ability estimates under various conditions. \textcolor{black}{Interestingly, when considering all respondents together, we observed that bias in $\theta$ estimates increased as the proportion of respondents with change-points decreased. This can be attributed to the estimation procedure: with fewer change-points in the sample, the model has less information to accurately estimate the change-point-related parameters, potentially affecting the overall accuracy of ability estimation. However, when examining only the subset of respondents with change-points, the bias shows the expected pattern of decreasing as the proportion of affected respondents decreases.}

This robust performance suggests that the model can reliably identify and account for changes in response behavior, even when these changes occur relatively late in the test. An important finding is the substantial reduction in bias of ability estimates when change-points are properly accounted for. This is evident in the simulation studies where the bias in $\theta$ estimates was significantly reduced after removing items affected by change-points. This improvement in estimation accuracy has important implications for the fairness and validity of test score interpretations.

The model demonstrates good performance in detecting the location of change-points, as evidenced by the relatively low MAE in change-point estimates across different simulation scenarios. This capability allows for a more precise identification of where in the test individual test-takers may begin to exhibit altered response behaviors. Analyses of the two educational testing datasets reveal meaningful patterns in change-point occurrences and their relationship to item characteristics and time pressure. These findings provide valuable insights into test-taking dynamics that were previously unobservable with standard IRT models.

Despite these promising results, there are limitations with the proposed framework and therefore scope for future research. One potential limitation is the assumption of a single change-point per individual. While this allows for tractable estimation and interpretation, it may not capture more complex patterns of behavior change that could occur in some testing scenarios. Future work could explore extensions to multiple change-points or more flexible change-point distributions. Another area for further investigation is the relationship between change-points and other observable variables, such as response times or demographic characteristics. Incorporating such covariates into the model could provide additional insights into the factors influencing changes in response behavior and potentially improve the accuracy of change-point detection.

The current model assumes local independence of item responses conditional on the latent trait and change-point. While this is a common assumption in IRT models, it may be violated in some testing contexts, particularly when items are clustered or when there are strong item position effects. Exploring relaxations of this assumption could lead to more flexible and realistic models of test-taking behavior. From a computational perspective, the estimation of our model, particularly when all parameters are unknown, can be computationally intensive. Future research could focus on developing more efficient estimation algorithms or exploring alternative inference approaches, such as variational methods or Markov Chain Monte Carlo techniques, to improve scalability to larger datasets or more complex model specifications.

The usefulness of the proposed methodology extends beyond the specific application to educational testing presented here. The ability to detect and account for individual-level change-points in latent variable models has potential applications in a wide range of fields where behavioral or cognitive processes may shift over time. For instance, in psychological assessment, our approach could be adapted to identify points at which individuals experience significant changes in their mental health or cognitive function during longitudinal studies. Marketing researchers could apply similar models to detect changes in consumer preferences or decision-making processes over time or across different contexts. In medical diagnosis, the model could be extended to analyze sequences of medical tests, helping to identify when a patient's condition may have significantly changed, potentially improving early detection of disease progression. The financial sector could benefit from this approach in credit scoring or fraud detection, helping to identify when an individual's financial behavior undergoes a significant shift, potentially indicating increased risk or fraudulent activity. Researchers in human-computer interaction could use similar models to detect changes in user behavior or engagement levels when interacting with software or online platforms.

From a statistical perspective, our work contributes to the literature on change-point detection in latent variable models. By demonstrating how change-point methods can be integrated with IRT models, we open up new possibilities for research at the intersection of these two fields. This approach could inspire similar developments in other areas of psychometrics or in the broader domain of latent variable modeling.

The practical implications of our model for educational testing are important. By providing a more accurate representation of test-taking behavior, including the effects of time pressure or fatigue, our approach can enhance the validity and fairness of test score interpretations. Test developers and psychometricians can use insights gained from this model to improve test design, potentially by adjusting test length, item ordering, time limits, or adaptive testing algorithms to mitigate the impact of behavioral changes on measurement accuracy. Moreover, the ability to identify individual-level change-points offers a new tool for understanding and addressing issues of test speededness. This could lead to more personalized testing experiences, where the pace or difficulty of items is dynamically adjusted based on real-time detection of changes in response behavior.

In conclusion, the change-point latent factor model presented in this paper represents an important advancement in the modeling of educational testing data. By explicitly accounting for individual-level changes in response behavior, our approach provides a more nuanced and accurate representation of test-taking processes. While there are areas for further refinement and extension, the model's performance in both simulated and real-data scenarios demonstrates its potential to enhance the precision and fairness of educational assessments. As the field of psychometrics continues to evolve, approaches that can capture the dynamic nature of cognitive processes and test-taking behaviors will become increasingly valuable. Our work offers a flexible framework that can be adapted and extended to address a wide range of challenges in educational measurement and beyond.

	\section*{Funding}
	Xiaoou Li's research is partially supported by NSF under the grant CAREER DMS-2143844.

	
	\clearpage
	\bibliography{bibliography}       
	
	
\end{document}